\documentclass[twocolumn]{aastex62}

\usepackage{graphicx}
\usepackage{amsmath}	
\usepackage{bm}
\usepackage{amssymb}
\usepackage{booktabs}
\usepackage{enumerate}
\usepackage{dirtytalk}
\usepackage{float}
\usepackage{enumitem}
\usepackage{booktabs}
\usepackage{lipsum} 
\usepackage{float}
\usepackage{xcolor}
\usepackage{multirow}
\usepackage{makecell}
\usepackage[english]{babel}

\renewcommand*{\thefootnote}{\fnsymbol{footnote}}

\accepted{March 2024, ApJ}

\submitjournal{AAS journals, October 2023}

\shorttitle{Model Uncertainty in Retrievals}
\shortauthors{Nixon et al.}

\begin{document}

\title{Methods for Incorporating Model Uncertainty into Exoplanet Atmospheric Analysis}

%A Framework for Marginalising over Model Uncertainty in Atmospheric Retrievals
%Methods for Incorporating Model Uncertainty into Atmospheric Retrievals

\author[0000-0001-8236-5553]{Matthew C. Nixon}
\email{mcnixon@umd.edu}
\affil{Department of Astronomy, University of Maryland, College Park, MD 20742, USA}

\author[0000-0003-0156-4564]{Luis Welbanks}
\affil{School of Earth and Space Exploration, Arizona State University, Tempe, AZ 85281, USA}
\affil{NHFP Sagan Fellow}

\author[0000-0002-1052-6749]{Peter McGill}
\affil{Department of Astronomy and Astrophysics, University of California, Santa Cruz, CA 95064, USA}
\affil{Space Science Institute, Lawrence Livermore National Laboratory, 7000 East Ave., Livermore, CA 94550, USA}

\author[0000-0002-1337-9051]{Eliza M.-R. Kempton}
\affil{Department of Astronomy, University of Maryland, College Park, MD 20742, USA}

\begin{abstract}
A key goal of exoplanet spectroscopy is to measure atmospheric properties, such as abundances of chemical species, in order to connect them to our understanding of atmospheric physics and planet formation. In this new era of high-quality JWST data, it is paramount that these measurement methods are robust. When comparing atmospheric models to observations, multiple candidate models may produce reasonable fits to the data. Typically, conclusions are reached by selecting the best-performing model according to some metric. This ignores model uncertainty in favour of specific model assumptions, potentially leading to measured atmospheric properties that are overconfident and/or incorrect. In this paper, we compare three ensemble methods for addressing model uncertainty by combining posterior distributions from multiple analyses: Bayesian model averaging, a variant of Bayesian model averaging using leave-one-out predictive densities, and stacking of predictive distributions. We demonstrate these methods by fitting the HST+Spitzer transmission spectrum of the hot Jupiter HD~209458b using models with different cloud and haze prescriptions. All of our ensemble methods lead to uncertainties on retrieved parameters that are larger, but more realistic, and consistent with physical and chemical expectations. Since they have not typically accounted for model uncertainty, uncertainties of retrieved parameters from HST spectra have likely been underreported. We recommend stacking as the most robust model combination method. Our methods can be used to combine results from independent retrieval codes, and from different models within one code. They are also widely applicable to other exoplanet analysis processes, such as combining results from different data reductions.
\end{abstract}


\section{Introduction}
\label{sec:intro}

Atmospheric characterisation of exoplanets has made rapid progress in recent years, as spectroscopic observations using both ground- and space-based facilities have yielded insights into a wide range atmospheric properties, including the presence and abundance of various chemical species, and information about the temperature structure and presence of clouds in the atmosphere \citep[see][for a review]{BarstowHeng2020}. In particular, initial observations from JWST have demonstrated sensitivity to a wide range of chemical species that were not previously accessible using space telescopes \citep[e.g.,][]{ERS2023,Bell2023}, demonstrating the potential for a revolution in our understanding of exoplanet atmospheres in the coming years.

In order to use these spectra to learn about atmospheric properties, we have to measure certain quantities of interest. One of the most common methods for making such measurements is to use atmospheric retrieval, i.e.\ inverse modelling of observed data using Bayesian sampling algorithms. In principle, this approach allows us to measure not only the values of parameters of interest, but also their uncertainties by exploring the full range of possible models that may be consistent with the observed data.

While this approach to model fitting allows us to explore the range of parameters that are consistent with observations within the confines of a single model, in reality there is a wide range of models that could be used for retrieval, and each of these may provide adequate descriptions of the observed data. Within a given retrieval framework, there are numerous variations of the forward model that can be considered. Take, for example, the temperature structure of the atmosphere, which is typically described using some kind of parametric pressure-temperature ($P$--$T$) profile. Past retrieval studies have used a single temperature to describe the atmosphere \citep[e.g.,][]{Tsiaras2018}, others have used some kind of one-dimensional $P$--$T$ profile \citep[e.g.,][]{Madhu2009,Guillot2010}, while others more recently have considered 2-D or 3-D profiles \citep[e.g.,][]{Welbanks2021_2D,Nixon2022}. A similarly large number of different approaches exist to model clouds and hazes, as discussed later in this work.

In addition to the range of possible forward models that can be implemented in a single code, a huge number of independent retrieval tools now exist, with $\sim$50 retrieval codes reported in the literature \citep{MacDonald2023}. While these codes in principle are solving the same radiative transfer equations to calculate spectra, it has been demonstrated that differences in their implementation do lead to non-negligible differences between forward models and retrieved parameter estimates \citep{Barstow2020_comparison}. It is often the case that when JWST spectra are published, multiple independent retrieval analyses are included \citep[e.g.,][]{Taylor2023,Coulombe2023}.

In both of the cases described above, it is not uncommon to find reasonable fits to the data with more than one candidate model. However, it is typically the case that ultimately a single model is chosen to give ``final'' reported results of the analysis, chosen by some measure such as maximising an information criterion \citep[e.g,][]{Sotzen2020,Spake2021}. By selecting a particular model, we ignore any existing uncertainty about the validity of the model in favour of the specific assumptions made by the chosen model. This could lead to overconfidence in reported measurements of atmospheric properties, weakening the reliability and credibility of any science that hinges on said measurements. While model uncertainty has been considered in many other disciplines of astrophysical modelling \citep[e.g.,][]{Gates1995,Rogers2010a,Moore2014,Dorn2015,Greco2016,Yu2018}, it is rarely considered in exoplanet atmospheric retrieval. It is therefore important, as our ability to observe exoplanet atmospheres makes unprecedented improvements, that we reconsider our approach to dealing with model uncertainty. To date, it has been common to marginalise over the parameters of a single atmospheric model. Here, we present a method to effectively marginalise over different models.

In this paper, we present several methods for combining the results of retrievals from multiple candidate models in a statistically sound manner. One of these methods (Bayesian model averaging) has been applied once before in a retrieval context \citep{Wakeford2017}, albeit with some differences to the method described here, while the other two have not been used for atmospheric retrievals. We describe the methods associated with each approach in Section \ref{sec:methods}. We present a case study of the atmosphere of the well-studied hot Jupiter HD~209458b in Section \ref{sec:results}, demonstrating how each method can be used to combine a range of different cloud and haze models into a single retrieval result. Finally, in Section \ref{sec:discussion} we explore the implications of our study for past retrieval analyses, and discuss the applicability of each method, alongside some practical points regarding their implementation.

\section{Approaches to Model Combination}
\label{sec:methods}

%Introduce Bayesian inference

Atmospheric retrieval frameworks typically carry out model fitting and parameter estimation using Bayesian inference, an application of Bayes' theorem to determine the probability distribution of a set of model parameters $\bm{\theta}$ given some data $\bm{d}$ and a model $M$:

\begin{align}
    p(\bm{\theta}|\bm{d},M) = \frac{p(\bm{d} | \bm{\theta},M) \, p(\bm{\theta} | M)}{p(\bm{d} | M)},
    \label{eq:bayes_thm}
\end{align}
where $p(\bm{\theta} | \bm{d}, M)$ is the posterior distribution (the probability distribution of model parameters given the data, i.e. the desired answer), $p(\bm{d} | \bm{\theta}, M)$ is the likelihood (the probability of the data given the model parameters), $p(\bm{\theta} | M)$ is the prior probability distribution of the model parameters independent of the data, and $p(\bm{d} | M)$ is the Bayesian evidence (the probability of the data given the model, independent of the parameters):

\begin{align}
    p(\bm{d} | M) = \int p(\bm{d} | \theta, M) \, p(\theta | M) \mathrm{d}\theta.
    \label{eq:evidence}
\end{align}

For exoplanet atmospheric retrieval, prior distributions are typically chosen so as to be as uninformative as possible, using uniform or log-uniform priors covering a wide range of possible values of each parameter, usually attempting to recover Jeffrey's priors. The likelihood function is then evaluated by sampling the prior space and evaluating how well a model generated with a given set of parameters fits the observed data. For data with independently distributed Gaussian errors (commonly assumed for exoplanet spectroscopy), the likelihood function is defined as

\begin{align}
    p \left( \bm{d}|\bm{\theta},M \right) = \prod_i^{N_{\rm obs}} \frac{1}{\sqrt{2 \pi} \sigma_i} \exp \left( -\frac{(d_i-\hat{d}_i)^2}{2\sigma_i^2} \right),
\end{align}
where $d_i$ is the observed value, $\sigma_i$ is the associated uncertainty, $\hat{d}_i$ is the model value of the $i^{\rm th}$ data point evaluated with parameters $\bm{\theta}$, and $N_{\rm obs}$ is the total number of observed data points.

Consider a scenario in which we fit our data set $\bm{d}$ using $K$ different models. These models may be combined by taking a weighted average using some vector of weights $\bm{W} \in \mathcal{S}^K$. The model combination problem involves finding an optimal set of weights $\hat{\bm{W}}$ such that the weighted sum of our $K$ posterior distributions is reflective of the posterior distribution marginalised over all considered models. $\mathcal{S}^K$ is defined as
\begin{align}
    \mathcal{S}^K = \bm{W} \in [0,1]^K : \sum_{k=1}^{K} W_k = 1.
\end{align}
The final posterior distribution resulting from this procedure is therefore
\begin{align}
    p (\bm{\theta}|\bm{d}) = \sum_{k=1}^{K} \hat{W}_k p \left( \bm{\theta}|\bm{d},M_k\right).
\end{align}

The remainder of this section describes three different methods for determining $\hat{\bm{W}}$.

\subsection{Bayesian model averaging} \label{subsec:bma}

According to Bayes' theorem, we can assign a probability to an individual model $M_j$ given the data:

\begin{align}
    p(M_j | \bm{d}) = \frac{ p(\bm{d} | M_j) \, p(M_j) }{\sum_{k=1}^K p(\bm{d} | M_k) \, p(M_k)}.
    \label{eq:model_prob}
\end{align}

We note that the quantity $p(\bm{d} | M_j)$ is the model evidence for model $M_j$ as defined in equation \ref{eq:evidence}. Equation \ref{eq:model_prob} can be rewritten in terms of Bayes factors, which are defined as the evidence ratio between two models:

\begin{align}
    B_{ij} = \frac{p \left( \bm{d} | M_i \right)}{p \left( \bm{d} | M_j \right)}.
\end{align}
By dividing out the evidence of an arbitrary reference model, $p \left( \bm{d} | M_1 \right)$, equation \ref{eq:model_prob} becomes

\begin{align}
    p(M_j | \bm{d}) = \frac{ B_{j1} \, p(M_j) }{\sum_{k=1}^K B_{k1} \, p(M_k)}.
\end{align}

Furthermore, if we assume that there is no \textit{a priori} basis for preferring a given model over the others (i.e., taking a uniform prior across models), this equation can be simplified to yield

\begin{align}
    p(M_j | \bm{d}) = \frac{ B_{j1} }{\sum_{k=1}^K B_{k1}}.
\end{align}

\noindent This provides us with a probability distribution over models, which we can use as weights for our combined posterior:

\begin{align}
    \hat{W}_j = \frac{ B_{j1} }{\sum_{k=1}^K B_{k1}}. \label{eq:bma}
\end{align}
This approach to model combination is known as Bayesian model averaging \citep[BMA,][]{Leamer1978}. This method requires calculation of the Bayesian evidence of each model under consideration. The most commonly employed sampling algorithm for atmospheric retrieval is Nested Sampling \citep{Skilling2006}. The Nested Sampling algorithm is specifically designed to calculate Bayesian evidences, making this application well-suited to combining atmospheric retrieval results.

A version of BMA has previously been applied to atmospheric retrievals \citep{Wakeford2017}, with some differences compared to the method described above. Since Markov chain Monte Carlo was used for parameter estimation rather than Nested Sampling, the Bayesian evidence of each model was not calculated. Instead, the authors found model weights using the Bayesian Information Criterion. Furthermore, rather than computing a weighted average of the full posterior, the weights were only used to compute an averaged mean and associated uncertainty for each parameter of interest. Here we seek a full combined posterior distribution rather than a single metric.

We note that the Bayesian evidence is strongly dependent on the choice of prior \citep{Trotta2008}, and so this method should ideally be applied only in the case where identical priors are specified, to avoid artificially inflating the evidence for a particular model. We discuss this in more detail, as well as possible ways to incorporate other sampling algorithms, in Section \ref{sec:discussion}.

Another drawback of BMA is that it can be biased by model expansion. Suppose that a data set has been fit using $K$ different models, each of which have similar Bayesian evidences. In this case, each model will be assigned a similar weight, $\hat{W}_k \sim 1/K$. Now suppose we add an extra model, $M_{K+1}$, which is in fact identical to one of the previous models, say model $M_1$. Since the evidences are unchanged, this would mean that model $M_1$ is given a higher weight ($\sim 2/(K+1)$) while the remaining models are given lower weights ($\sim 1/(K+1)$). While in practice, it is unlikely that two identical models would be included in one analysis, the problem could arise if a number of models in the set are very similar to each other. In atmospheric retrieval, an example of this could be where multiple independent teams conduct retrievals of a spectrum, but two or more of those teams use the same retrieval framework. This issue can be avoided by using stacking of predictive distributions (see Section \ref{subsec:stacking}).

\subsection{Pseudo Bayesian model averaging} \label{subsec:pseudo_bma}

Bayesian evidence is not the only quantity that has been proposed as a weighting method for combining inferences from multiple models. As an alternative, \citet{Geisser1979} proposed the use of ``pseudo Bayes factors'' constructed by taking the product of Bayesian leave-one-out (LOO) predictive densities. For a given model $M$, this quantity is the exponent of the expected log posterior predictive density of the model, elpd$_{\mathrm{LOO},M}$, which is defined as follows:

\begin{align}
    \mathrm{elpd}_{\mathrm{LOO},M} &= \sum_{i=1}^{N_{\mathrm{obs}}} \mathrm{elpd}_{\mathrm{LOO},M, i} \nonumber \\
    &= \sum_{i=1}^{N_{\mathrm{obs}}} \log p \left( d_i | \bm{d}_{-i}, M \right),
\end{align}

\noindent where $p \left( d_i | \bm{d}_{-i}, M \right)$ is the posterior predictive density of data point $d_i$ under $M$ trained on the remainder of the data set after $d_i$ has been removed, which we label $\bm{d}_{-i}$. The value of elpd$_{\mathrm{LOO},M}$ reflects the out-of-sample predictive accuracy of the model trained on the entire observed data set, with higher values indicating better predictive performance. It is asymptotically equal to the widely applicable information criterion \citep[WAIC,][]{Watanabe2010}. However, elpd$_{\mathrm{LOO},M}$ is more robust than WAIC, particularly in cases with weak priors \citep{Vehtari2015}, as is often the case in exoplanet atmospheric retrieval.

We can compute $p \left( d_i | \bm{d}_{-i}, M \right)$ by taking the expectation of the likelihood of $d_i$ with respect to the
posterior of the model fitted to $\bm{d}_{-i}$:

\begin{align}
    p \left( d_i | \bm{d}_{-i}, M \right) = \int p \left( d_i | \bm{\theta}, M \right) p \left( \bm{\theta} | \bm{d}_{-i}, M \right) \mathrm{d} \bf{\theta}.
\end{align}

\noindent Calculating $p \left( d_i | \bm{d}_{-i}, M \right)$ would require refitting the model $N_{\rm obs}$ times. This can be circumvented by using Pareto smoothed importance sampling \citep[PSIS,][]{Vehtari2017} to approximate the LOO predictive densities. PSIS-LOO is starting to be applied to problems in various areas of astrophysics \citep[e.g.,][]{Morris2021,Meier2022, McGill2023, Neil2022, Challener2023} and further details on its current application to atmospheric retrieval can be found in \citet{Welbanks2023}. PSIS-LOO works by drawing $S_p$ samples from the posterior fit to the full data set, and for each sample $s$ computing

\begin{align}
    r^s_i = \frac{1}{p \left( d_i|\bm{\theta}^s, M \right)} \propto \frac{p \left( \bm{\theta}^s | \bm{d}_{-i}, M \right)}{p \left( \bm{\theta}^s | \bm{d}, M \right)}.
\end{align}

The quantity $r^s_i$ is the importance weight for data point $i$ and sample $s$. The raw importance weights may be dominated by extreme values, making the importance sampling unreliable \citep{Vehtari2015}. To resolve this problem, we fit the distribution of importance weights with a Pareto distribution \citep{Zhang2009}, and draw smoothed importance weights $w^s_i$ which replace $r^s_i$ \citep[see][for details]{Vehtari2015}. The expression for LOO predictive density then becomes

\begin{align}
    p \left( d_i | \bm{d}_{-i}, M \right) &= \int p \left(d_i | \bm{\theta}, M \right) \frac{p \left( \bm{\theta} | \bm{d}_{-i}, M \right)}{p \left( \bm{\theta} | \bm{d}, M \right)} p \left( \bm{\theta} | \bm{d}, M \right) \mathrm{d} \bm{\theta} \nonumber \\
    & \approx \frac{\sum_{s=1}^{S_p} w_i^s p \left( d_i | \bm{\theta}^s, M \right) }{\sum_{s=1}^{S_p} w_i^s p}.
\end{align}

One approach to model weighting using $\rm{elpd}_{\rm LOO}$ would be to choose weights

\begin{align}
    \hat{W}_k = \frac{\exp \left( \mathrm{elpd}_{\mathrm{LOO},M_k} \right)}{\sum_{k=1}^{K} \exp \left( \mathrm{elpd}_{\mathrm{LOO},M_k}\right)}.
\end{align}
However, this approach does not account for the uncertainty in $\mathrm{elpd}_{\mathrm{LOO},M}$ due to the finite number of available samples. We can do so by calculating the standard error,
\begin{align}
    \mathrm{SE} \left( \mathrm{elpd}_{\mathrm{LOO},M} \right) &= \sqrt{N_{\mathrm{obs}} \mathrm{Var} \left( \mathrm{elpd}_{\mathrm{LOO},M} \right) } \nonumber \\
    & = \sqrt{ \sum_{i=1}^{N_{\mathrm{obs}}} \left( \mathrm{elpd}_{\mathrm{LOO},M,i} - \frac{\mathrm{elpd}_{\mathrm{LOO},M}}{N_{\mathrm{obs}}} \right)^2 },
\end{align}
and modifying our weights using the log-normal approximation, yielding
\begin{align}
    \hat{W}_k = \frac{\exp \left[ \mathrm{elpd}_{\mathrm{LOO},M_k} - \frac{1}{2} \mathrm{SE} \left( \mathrm{elpd}_{\mathrm{LOO},M_k} \right) \right]}{\sum_{k=1}^{K} \exp \left[ \mathrm{elpd}_{\mathrm{LOO},M_k} - \frac{1}{2} \mathrm{SE} \left( \mathrm{elpd}_{\mathrm{LOO},M_k} \right) \right]}. \label{eq:pseudo-bma}
\end{align}
Following \citet{Yao2018}, we call this method pseudo BMA. 

Pseudo BMA has a number of advantages over BMA. The chosen weights in pseudo BMA are calculated by drawing samples from the posterior of the trained models. While this still depends on the prior, these weights are much less likely to be biased by prior choice unless said choice excludes the high-likelihood region of the parameter space (see Section \ref{sec:discussion}). Furthermore, unlike in BMA where uncertainty in the Bayesian evidence is not accounted for, the uncertainty in $\mathrm{elpd}_{\mathrm{LOO},M}$ is directly incorporated, which has the effect of regularising the weights \citep{Yao2018}. However, pseudo BMA can still be biased by model expansion, as described in Section \ref{subsec:bma}.

\subsection{Stacking of predictive distributions} \label{subsec:stacking}

Stacking of predictive distributions relies on defining a scoring rule $s(P,\bm{d})$, which is a function of some probability distribution $P$ and the outcome $\bm{d}$ of a data generator, which we assume is drawn from another probability distribution $Q$. For a given scoring rule, we can define a divergence $D$ between distributions $P$ and $Q$:

\begin{align}
    D(P,Q) = S(Q,Q) - S(P,Q),
\end{align}
where
\begin{align}
    S(P,Q) = \int s(P,\bm{d}) \mathrm{d} Q(\bm{d}).
\end{align}

For the problem considered here, $P$ is the posterior distribution obtained by fitting a model (or combination of models) to a data set (spectrum) $\bm{d}$. The scoring rule encapsulates how well $P$ can generate predictions that are close to our observed outcome, $d$. Therefore, $s(P,\bm{d})$ is large (and $D(P,Q)$ is small) when $P$ is well-chosen relative to $\bm{d}$. Following \citet{Yao2018}, we take $D$ to be the Kullback-Leibler divergence of $P$ from $Q$:

\begin{align}
    D(P,Q) &= \Delta_{\rm KL}(Q,P) \nonumber \\
    &= \int^{\infty}_{-\infty} q(x) \log \frac{q(x)}{p(x)} \mathrm{d} x \\
    \implies s(P,\bm{d}) &= \log p(\bm{d}), \label{eq:logscore}
\end{align}
where $p$ and $q$ are the probability densities of $P$ and $Q$. The goal of stacking is to find a weighted average of our $K$ posterior probability distributions that minimises $D$, or equivalently, maximises $s$. Using the same definitions of $\bm{W}$ and $\mathcal{S}^K$ as before, this can be written as

\begin{align}
    \min_{\bm{W} \in \mathcal{S}^K} D \left( \sum_{k=1}^K W_k p\left(\tilde{\bm{d}}|\bm{d},M_k\right), q\left(\tilde{\bm{d}}|\bm{d}\right) \right),
\end{align}
or
\begin{align}
    \max_{\bm{W} \in \mathcal{S}^K} s \left( \sum_{k=1}^K W_k p\left(\tilde{\bm{d}}|\bm{d},M_k\right), \bm{d} \right),
    \label{eq:max_score}
\end{align}
where $p(\tilde{\bm{d}}|\bm{d},M_k)$ is the predictive density of new data $\tilde{\bm{d}}$ under model $M_k$ trained on the observed data $\bm{d}$, and $q(\tilde{\bm{d}}|\bm{d})$ is the density of the true data generating distribution. Since we lack the full predictive distribution $p(\tilde{\bm{d}}|\bm{d},M_k)$ for any new data, we can instead approximate this distribution with the LOO predictive distribution $p(d_i|\bm{d}_{-i}, M_k)$. Replacing the corresponding term in equation \ref{eq:max_score} and using the scoring rule from equation \ref{eq:logscore}, the equation for the optimal model weights becomes
\begin{align}
    \hat{\bm{W}} = \max_{\bm{W} \in \mathcal{S}^K} \left[ \frac{1}{n} \sum_{i=1}^{N_{\rm obs}} \log \sum_{k=1}^K W_k p\left(d_i|\bm{d}_{-i},M_k\right) \right]. \label{eq:stacking}
\end{align}

We can approximate $p\left(d_i|\bm{d}_{-i},M_k\right) = \exp \left( \mathrm{elpd}_{i,M_k} \right)$ using the methods described in Section \ref{subsec:pseudo_bma}. Therefore, to compute this quantity for a given set of weights $\bm{W}$, we first compute $\mathrm{elpd}_{i,M_k}$ for each data point in the spectrum and for each model under consideration. Then we calculate the weighted sum of this value for each model. Finally, we sum over each data point in the model. This gives an overall measure of the predictive performance of the weighted average model.

Finding the appropriate weights $\hat{\bm{W}}$ out of all possible weights $\bm{W} \in \mathcal{S}^K$ therefore becomes an optimization problem, meaning standard optimization algorithms can be used to search for the optimal set of weights. In this work we use the Sequential Least Squares Programming algorithm to find the optimal weights \citep{Kraft1988}.

Stacking finds the combination of predictive distributions that is closest to the data generating process with respect to the chosen scoring rule. In other words, we are considering all possible posterior distributions that could be constructed using a weighted average of our individual posteriors, and finding the combination which has the highest likelihood of having generated the observed spectrum. Since stacking only depends on the space covered by all models considered, it does not suffer from bias due to model expansion \citep{Yao2018}.

%%%%%%%%%%%%%%%%%%%%%%%%%

\begin{table}
    \centering
    \setlength{\arrayrulewidth}{1.3pt}
    \begin{tabular}{cc}
    	\hline
		Parameter & Prior \\
		\hline
        \multicolumn{2}{c}{Common parameters} \\
        \hline
        $\log_{10} X_{\rm i}$ & [$-12,0$] \\
        $T_0$ (K) & [$800,1550$] \\
        $\alpha_1,\alpha_2$ $(\mathrm{K}^{-\frac{1}{2}})$ & [0.02,2.0] \\
        $\log_{10} P_1, \log_{10} P_2, \log_{10} P_{\rm ref}$ (bar) & [-6,2] \\
        $\log_{10} P_3$ (bar) & [-2,2]  \\
        \hline
        \multicolumn{2}{c}{PL+G parameters} \\
        \hline
        $\log_{10} a$ & [-4,10] \\
        $\gamma$ & [-20,2] \\
        $\log_{10} P_c$ (bar) & [-6,2] \\
        \hline
        \multicolumn{2}{c}{A\&M parameters} \\
        \hline
        $\log_{10} P_{\rm base}$ (bar) & [-6,2] \\
        $\log_{10} f_{\rm cond}$ & [-10,0] \\
        $\log_{10} K_{zz}$ (cm$^2\,$s$^{-1}$) & [6.5,10.5] \\
        $f_{\rm sed}$ & [1,5] \\
        \hline
        \multicolumn{2}{c}{Patchy cloud parameters} \\
        \hline
        $\phi_{\rm c},\phi_{\rm h},\phi_{\rm c+h}$ & [0,1] \\
        \hline
    \end{tabular}
    \caption{Priors for all retrievals presented in this work. Parameters shared between all models are shown at the top, followed by parameters specific to the power-law haze + grey cloud (PL+G) models, then the Ackerman \& Marley (A\&M) models, and finally the patchy cloud models. All priors are uniform within the ranges shown (or log-uniform as specified). Quantities without specified units are dimensionless. $X_i$ denotes the volume mixing radio of chemical species $i$. If the sum of the mixing ratios of the chemical species is greater than one, the likelihood is automatically set to zero to prevent unphysical solutions.}
    \label{tab:priors}
\end{table}

\section{Case Study: the cloudy atmosphere of HD~209458~b}
\label{sec:results} 

In order to demonstrate the methods for incorporating model uncertainty described in Section \ref{sec:methods}, we conduct a series of retrievals of the transmission spectrum of the hot Jupiter HD~209458~b \citep{Charbonneau2000}, which consists of observations taken using the Hubble Space Telescope (HST) with both the Space Telescope Imaging Spectrograph (STIS) and Wide Field Camera 3 (WFC3) instruments, as well as additional observations from the Spitzer Space Telescope \citep{Sing2016}. For this case study, we consider a set of models in which we vary the methodology used to account for clouds and hazes, including a wide range of approaches that have been presented in the literature. We produce weighted averages of the full set of models using each of the methods described above. A full list of parameters and associated priors is provided in Table \ref{tab:priors}.

 We choose this set of models since clouds and hazes are extremely important to consider when modelling transmission spectra, and there is no consensus approach to implementing cloud and haze models that has been adopted across all retrieval frameworks. More importantly, previous studies of this planet have resulted in widely different abundance estimates, largely influenced by their treatment of clouds and hazes. For instance, \cite{Madhusudhan2014} inferred a significantly sub-solar H$_2$O volume mixing ratio using cloud-free models, while recent reanalyses have resulted in abundances consistent with solar and super-solar values \citep[e.g.,][]{Welbanks2019_mz}. Therefore, in the presence of multiple models with differing inferences, in this work we explore the resulting constraints after marginalising over a set of atmospheric models which use different aerosol prescriptions. 
 
 Our approaches to model combination can readily be applied to other sets of models: for example, a set of retrievals of an emission spectrum could consider models with a range of parametric temperature profiles, since emission spectra are particularly sensitive to thermal structure. Our methods could also be applied to models with the same parameters, but different implementations of radiative transfer (i.e.\ different retrieval codes). Finally, we anticipate this approach to be helpful in future searches for multidimensional effects in transmission spectra, emission spectra, and phase curves, where several modelling approaches exist and where considering multidimensional effects may result in different inferred atmospheric properties \citep[e.g.,][]{Nixon2022}. 

\subsection{Atmospheric retrieval setup}

We use the \textit{Aurora} retrieval framework \citep{Welbanks2021}, which combines an atmospheric forward model with a Bayesian parameter estimation scheme. The implementation of \textit{Aurora} used in this work incorporates a number of features from the \textsc{Aura} framework \citep{Pinhas2018,Welbanks2019,Nixon2020,Nixon2022}.

\textit{Aurora} calculates radiative transfer in the atmosphere of an exoplanet in hydrostatic equilibrium transiting its host star, assuming plane-parallel geometry. The forward model incorporates a parametric pressure-temperature ($P$--$T$) profile, absorption from a range of chemical species, including collision-induced absorption (CIA), and a number of parametric treatments of cloud and haze opacity. These inputs are used to calculate a transmission or emission spectrum at high resolution, which is convolved with the point spread function of the relevant instrument and binned to the resolution of the observations in order to be compared with real data.

For the retrievals shown in this work, we assume a H/He-dominated atmosphere in transmission geometry with solar relative abundances of H and He \citep{Asplund2009}.  We include CIA due to H$_2$-H$_2$ and H$_2$-He \citep{Richard2012}. We also include absorption due to the following chemical species: Na \citep{Allard2016}, K \citep{Allard2019}, H$_2$O \citep{Rothman2010}, CH$_4$ \citep{Yurchenko2014}, NH$_3$ \citep{Yurchenko2011}, HCN \citep{Barber2014}, CO and CO$_2$ \citep{Rothman2010}. Volume mixing ratios ($X_i$) for all species other than H and He are included as free parameters, with the remaining component of the atmosphere assumed to be H/He in solar proportions \citep{Asplund2009}.

For all retrievals shown here, we adopt the parametric $P$--$T$ profile from \citet{Madhu2009}, which divides the atmosphere into three layers. This prescription consists of six free parameters: the temperature at the top of the atmosphere, assumed to be 1 $\mu$bar ($T_0$), the boundary in pressure between the first and second layers ($P_1$), a parameter that allows for a thermal inversion in the second layer ($P_2$, where $P_2>P_1$ leads to a thermal inversion), the boundary in pressure between the second and third layers ($P_3$), and two parameters which describe the curvature of the temperature profile in the upper two layers ($\alpha_1,\alpha_2$). Note that since we do not expect thermal inversions at the atmospheric terminator, we restrict $P_2 \leq P_1$ for retrievals of transmission spectra. The third (deepest) layer is assumed to be isothermal. This $P$--$T$ profile is widely adopted among retrieval codes \citep[e.g.,][]{Line2013,Pinhas2018,Blecic2022} and has been extensively validated against temperature structures from self-consistent atmospheric models and solar system observations.

Parameter estimation is carried out using Nested Sampling \citep{Skilling2006}. Specifically, we use PyMultiNest \citep{Buchner2014}, a Python interface for MultiNest \citep{Feroz2008,Feroz2009}. As discussed in Section \ref{sec:intro}, the use of Nested Sampling allows us to calculate the Bayesian evidence of each model under consideration, enabling the computation or relative Bayes factors for BMA.

\subsubsection{Clouds and hazes} \label{subsubsec:clouds}

The presence of clouds and photochemical hazes in an atmosphere can strongly impact and often inhibit spectroscopic observations \citep{Barstow2020_clouds}. The observational signatures of clouds and hazes depend on the wavelengths being observed: at visible wavelengths, aerosols may lead to a steep downward slope in transmission spectra \citep{Pont2013} or reflected stellar spectra in emission spectra \citep{Marley2013}, whereas in the infrared, clouds and hazes lead to muted spectral features in both transmission and emission spectra, with the effect being particularly prominent for transmission spectra \citep{Fortney2005}. Aerosol species also possess some features in the far infrared \citep{Pinhas2017}. Clouds and hazes have been proposed to explain atmospheric observations of numerous exoplanets. For example, \citet{Sing2016} invoked the presence of clouds in several hot Jupiters to explain muted H$_2$O features compared to what would be expected from chemical equilibrium with solar elemental abundances, while clouds and photochemical hazes have been used to explain the flat transmission spectrum of the sub-Neptune GJ~1214~b \citep{Kreidberg2014_GJ,Gao2023}. Degeneracies between aerosol parameters and other atmospheric parameters have been widely reported \citep[e.g.,][]{Benneke2012,Line2016,Welbanks2019}, with cloudy atmospheres appearing similar to those with depleted chemical abundances \citep{Barstow2017,Pinhas2019} or high mean molecular weight \citep[e.g.,][]{Libby-Roberts2022}.

Due to the apparent ubiquity of clouds and hazes in exoplanet atmospheres, and their significant effect on observed spectra, it is necessary to consider cloud and haze opacity in any retrieval framework. While sophisticated cloud and haze microphysical models that incorporate formation,
growth, sedimentation, and mixing in a self-consistent manner have been developed \citep[e.g.,][]{Helling2008,Lavvas2017,Gao2018}, these models are too computationally intensive for incorporation into retrieval forward models. Instead, simpler parametric models are used that can reproduce the observable effects of clouds and hazes while remaining fast to compute \citep[e.g.,][]{Ackerman2001,LDE2008}. A number of these parametric models exist, and there is no strong consensus as to which specific model to apply in a retrieval context. Furthermore, the choice of cloud model can affect retrieved posteriors and parameter estimates of other properties such as the temperature structure and abundances of chemical species to varying degrees \citep{Mai2019,Barstow2020_clouds}. 

Since the problem of accounting for clouds and hazes in a retrieval involves consideration of multiple competing models, we use it as a demonstration of the model combination methods described in Section \ref{sec:methods}. We consider the following cloud and haze prescriptions:

\textit{1. Cloud-free model.} This is the simplest model considered, where we assume a cloud and haze-free atmosphere. No changes are required to the forward model.

\textit{2. Power-law haze + grey cloud model.} In this model, a grey cloud deck is added by setting the optical depth to infinity for all pressures larger than the cloud-top pressure $P_{\rm cloud}$. Hazes are modelled as deviations from H$_2$ Rayleigh scattering \citep{LDE2008}, with a cross-section

\begin{align}
    \sigma_{\lambda, \rm haze} = a \sigma_0 \left( \dfrac{\lambda}{\lambda_0} \right)^{\gamma},
    \label{eq:haze}
\end{align}

\noindent where $a$ is the Rayleigh enhancement factor, i.e. the amplitude of the haze cross-section relative to H$_2$ Rayleigh scattering, $\gamma$ is the scattering slope, and $\sigma_0 = 5.31 \times 10^{-31}\,$m$^2$ is the cross-section due to H$_2$ Rayleigh scattering at a reference wavelength $\lambda_0 = 3.5 \times 10^{-7}\,$m \citep{Dalgarno1962}. In total, this adds three free parameters to the forward model: $a, \gamma$ and $P_{\rm cloud}$.

\textit{3. Two-sector power-law haze + grey cloud model.} This model accounts for clouds and hazes in the same manner as above, but also allows for the possibility of non-uniform cloud coverage \citep{Line2016} through the parameter $\phi_{\rm{c+h}}$, which sets the fraction of the atmospheric terminator that is covered by clouds and hazes. Therefore, this prescription adds four free parameters to the forward model: $a, \gamma, P_{\rm cloud}$ and $\phi_{\rm{c+h}}$. This prescription for clouds and hazes has been widely implemented in atmospheric retrievals \citep[e.g.][]{MacDonald2017,vonEssen2020,Taylor2023}.

\textit{4. Three-sector power-law haze + grey cloud model.} This is a variation on the two-sector model, also using equation \ref{eq:haze} to compute haze cross-sections and including an opaque cloud deck at $P_{\rm cloud}$. However, rather than splitting the terminator into a clear sector and one that contains both clouds and hazes, this model separates the cloudy and hazy sectors of the atmosphere, resulting in three sectors controlled by two parameters: $\phi_{\rm{c}}$ and $\phi_{\rm{h}}$. This adds an additional free parameter to the forward model, leading to five parameters in total. This model was first presented in \citet{Welbanks2021}, alongside the following model.

\textit{5. Four-sector power-law haze + grey cloud model.} This model is the same as above, except it splits the atmosphere into four sectors: a cloud/haze free sector, a sector with a grey cloud deck, a sector with hazes, and a sector with both clouds and hazes. It uses all three of the parameters $\phi_{\rm{c}},\phi_{\rm{h}}$ and $\phi_{\rm{c+h}}$, resulting in six free parameters in total.

\textit{6. Ackerman \& Marley model.} Unlike the previous models, this prescription incorporates Mie scattering from a condensate species following a method first proposed by \citet{Ackerman2001}. For the purposes of this study, we assume a condensate composition of pure MgSiO$_3$ (enstatite), which has previously been proposed to explain observed haze scattering slopes of a hot Jupiter \citep{LDE2008}. We adopt the cross-sections presented in \citet{Pinhas2017}\footnote{\texttt{github.com/Exo-worlds/Mie\_data}}. In this model, cloud profiles are calculated by balancing the downward sedimentation of condensate particles and the upward vertical mixing of condensate particles and vapour:
\begin{align}
    \frac{\partial q_c}{\partial z} + \frac{\partial q_v}{\partial z} + \frac{v_{\rm sed} q_c}{K_{zz}} = 0,
\end{align}
where $q_c$ and $q_v$ are the mass mixing ratios of condensate particles and vapour, $v_{\rm sed}$ is the sedimentation velocity of condensate particles, and $K_{zz}$ is the eddy diffusion coefficient, which describes the level of vertical mixing due to convection or turbulence. The condensate mixing ratio takes the form
\begin{align}
    q_c = 
    \begin{cases}
    f_{\rm cond} \left( P/{P_{\rm base}} \right)^{f_{\rm sed}}, & P \leq P_{\rm base}, \\
    0, & P > P_{\rm base},
    \end{cases}
\end{align}
where $P_{\rm{base}}$ is the pressure at the cloud base, $f_{\rm cond}$ is the condensate mixing ratio at the cloud base, and $f_{\rm{sed}} = v_{\rm sed}/v^*$ is the ratio of the particle sedimentation velocity to the characteristic vertical mixing velocity. This characteristic velocity $v^* = K_{zz}/H_{\rm sc}$, where $H_{\rm sc} = k_BT/\mu_g g$ is the atmospheric scale height and  $\mu_g$ is the mean molecular weight of the background gas.

This model has been implemented in \textit{Aurora} for the first time for the purpose of this study, with full details of the calculations provided in the Appendix. Four free parameters are required: $K_{zz}$, $f_{\rm cond}$, $f_{\rm{sed}}$ and $P_{\rm{base}}$.

\textit{7. Two-sector Ackerman \& Marley model.} This is the same model as above, but with an additional free parameter to account for inhomogeneous cloud cover. Similarly to the two-sector power-law haze + grey cloud model, we use an additional free parameter $\phi_{\rm c}$ which represents the fraction of the terminator that is covered by these condensates.

\subsection{Retrievals of HD~209458~b} \label{subsec:hd209}

\begin{figure*}
\centering
\includegraphics[width=\linewidth,trim={0 0 0 0},clip]{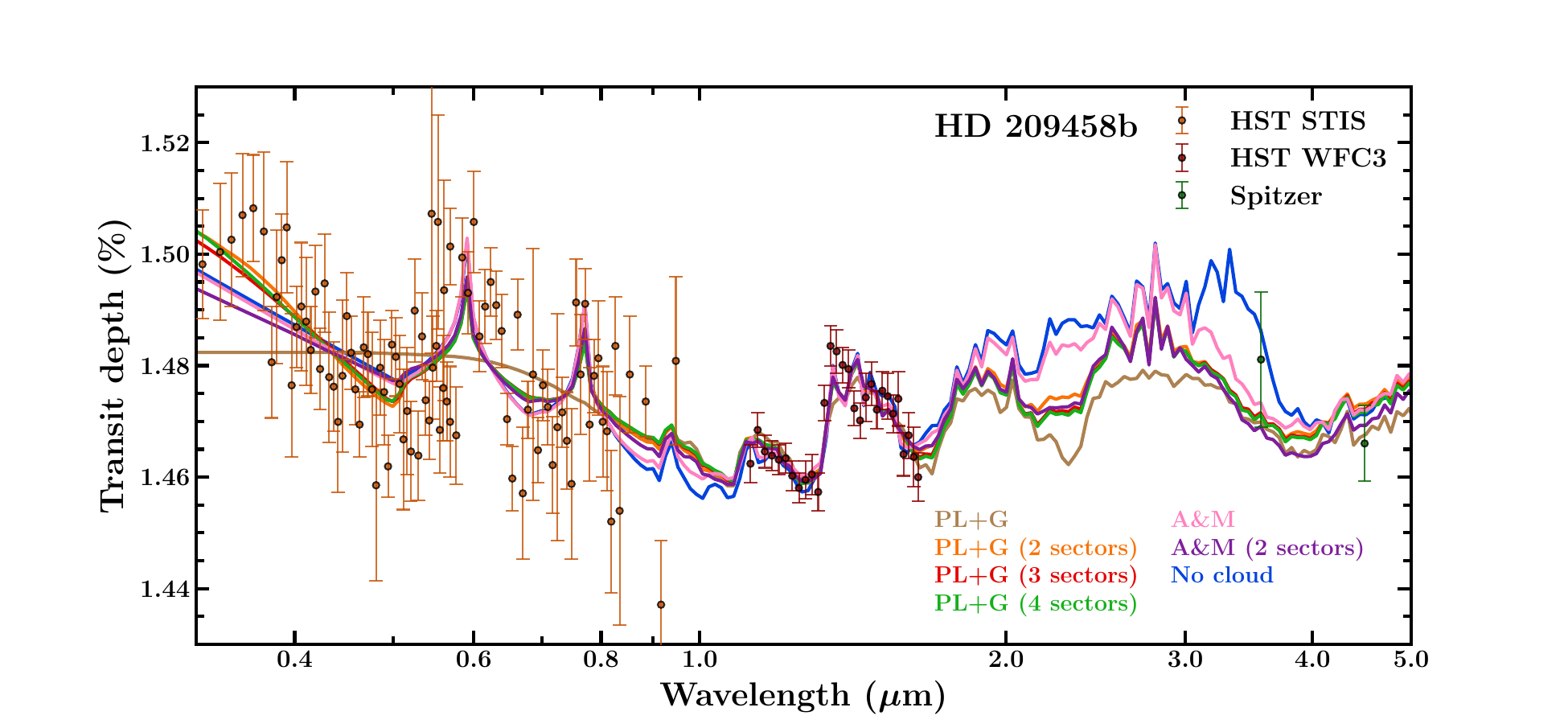}
    \caption{Transmission spectrum of HD~209458b, observed with HST STIS (orange points), HST WFC3 (red points), and Spitzer (green points). Overplotted are median model spectra drawn from the posteriors of each of the seven retrievals described in Section \ref{subsec:hd209}. All models are the same except for their treatment of clouds and hazes. Median forward models have been downsampled to a resolution $R=200$ for clarity. PL+G refers to power-law haze + grey cloud models, and A\&M refers to Ackerman \& Marley cloud models.}
    \label{fig:all_medians}
\end{figure*}

\begin{table*}
    \centering
    \setlength{\arrayrulewidth}{1.3pt}
    \begin{tabular}{ccccccc}
    	\hline
		Model & Retrieved $\log_{10} X_{\rm{H_2O}}$ & $\log \mathcal{Z}$ & BMA wt.\ (\%) & $\mathrm{elpd_{LOO}}$ (SE) & Pseudo BMA wt.\ (\%) & Stacking wt.\ (\%) \\
		\hline
        PL+G             & $-0.53^{+0.09}_{-0.07}$ & 935.29 &  0.00 & 947.23 (12.21) &  0.00 &  0.00 \\
        PL+G (2 sectors) & $-4.53^{+0.34}_{-0.30}$ & 957.67 & 21.66 & 968.70 (11.16) &  9.93 & 10.06 \\
        PL+G (3 sectors) & $-4.06^{+0.75}_{-0.52}$ & 958.08 & 31.50 & 968.78 (11.29) & 10.07 &  9.91 \\
        PL+G (4 sectors) & $-3.89^{+0.78}_{-0.63}$ & 958.40 & 45.20 & 970.74 (11.09) & 79.28 & 64.72 \\
        A\&M             & $-4.80^{+0.14}_{-0.15}$ & 951.09 &  0.03 & 964.36 (11.88) &  0.09 &  4.56 \\
        A\&M (2 sectors) & $-4.92^{+0.16}_{-0.17}$ & 955.07 &  1.61 & 966.35 (11.96) &  0.63 &  7.39 \\
        Cloud-free       & $-4.98^{+0.14}_{-0.15}$ & 949.19 &  0.00 & 959.85 (12.85) &  0.00 &  3.36\\
        \hline
    \end{tabular}
    \caption{Retrieved H$_2$O abundances alongside Bayesian evidences ($\log \mathcal{Z}$), $\rm elpd_{LOO}$ scores (with standard errors), and weights (wt.) for each of the models described in Section \ref{subsec:hd209} produced by all three model combination methods: Bayesian model averaging (BMA, see equation \ref{eq:bma}), pseudo-BMA (see equation \ref{eq:pseudo-bma}), and stacking (see equation \ref{eq:stacking}). PL+G refers to power-law haze + grey cloud models, and A\&M refers to Ackerman \& Marley cloud models.}
    \label{tab:z_elpd}
\end{table*}

We conduct retrievals of the transmission spectrum of HD~209458~b using each of the seven models described above. The median spectral fit to the data drawn from the posteriors of each model is shown in Figure \ref{fig:all_medians}. We find that all but one of the candidate models are able to find a reasonable fit to the data. The $\chi^2_{\rm{red}}$ for the best-fit models ranges from 1.61$-$1.73 across the range of models considered, with the exception of the single-sector power-law haze + grey cloud model, for which $\chi^2_{\rm{red}}=2.19$. We see very similar performance in the HST WFC3 band in particular. However, the multi-sector power-law haze + grey cloud models find a somewhat better fit to the blue end of the STIS data compared to the cloud-free and Ackerman \& Marley models. This is reflected in the Bayesian evidences and $\mathrm{elpd_{LOO}}$ scores, which are higher for the models which use a power-law haze + grey cloud prescription (see Table \ref{tab:z_elpd}).

\begin{figure*}
\centering
\includegraphics[width=0.45\linewidth]{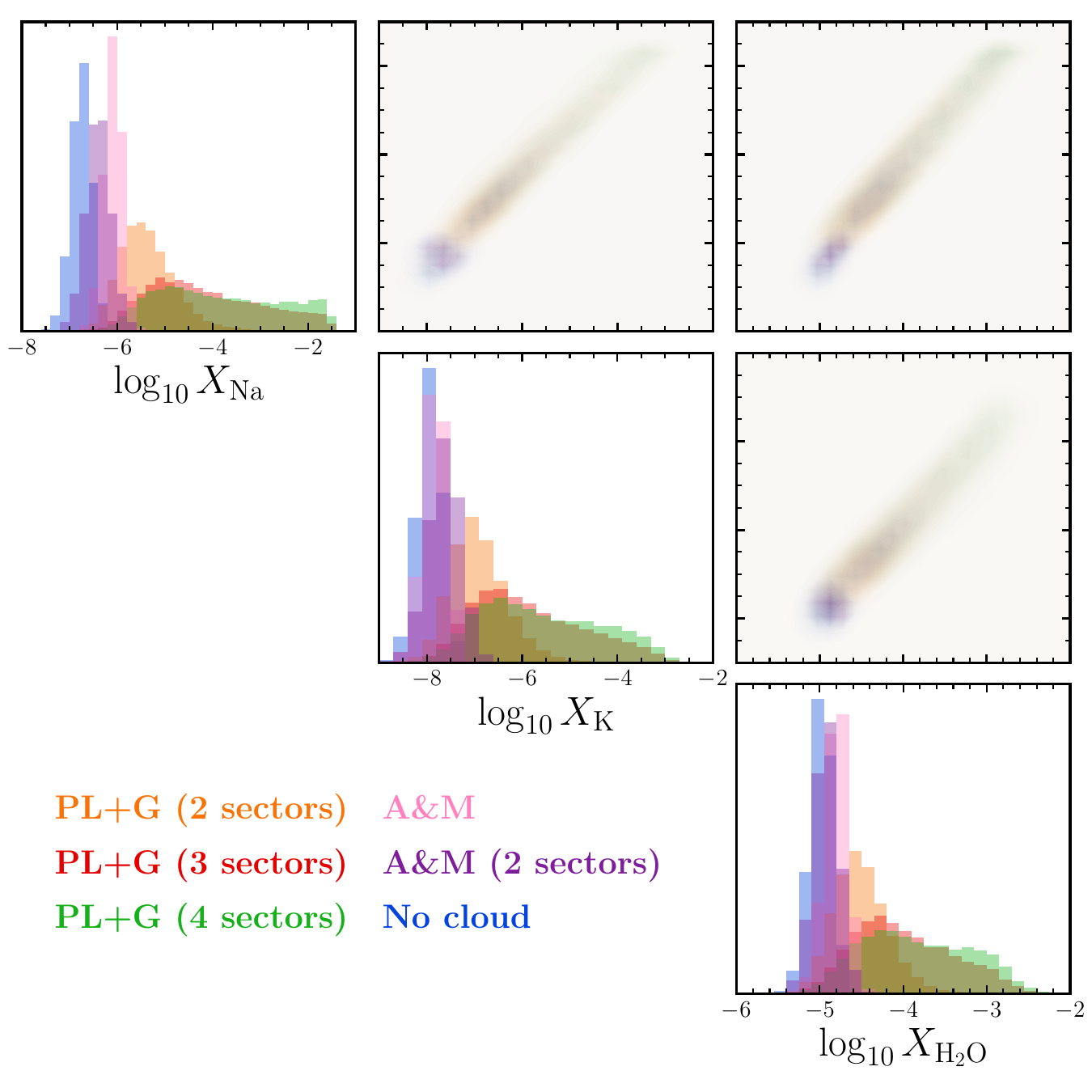}
\includegraphics[width=0.45\linewidth]{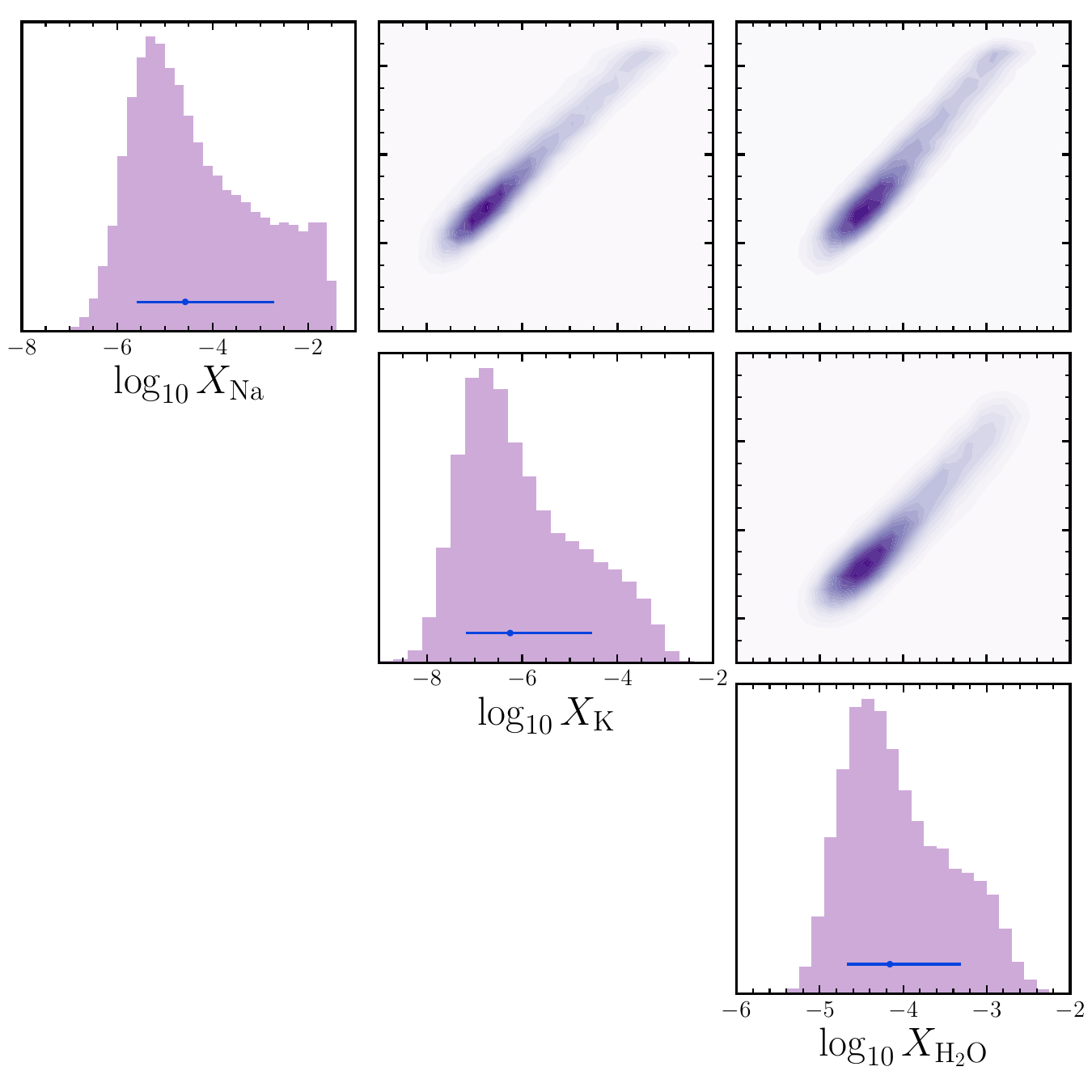}

\includegraphics[width=0.45\linewidth]{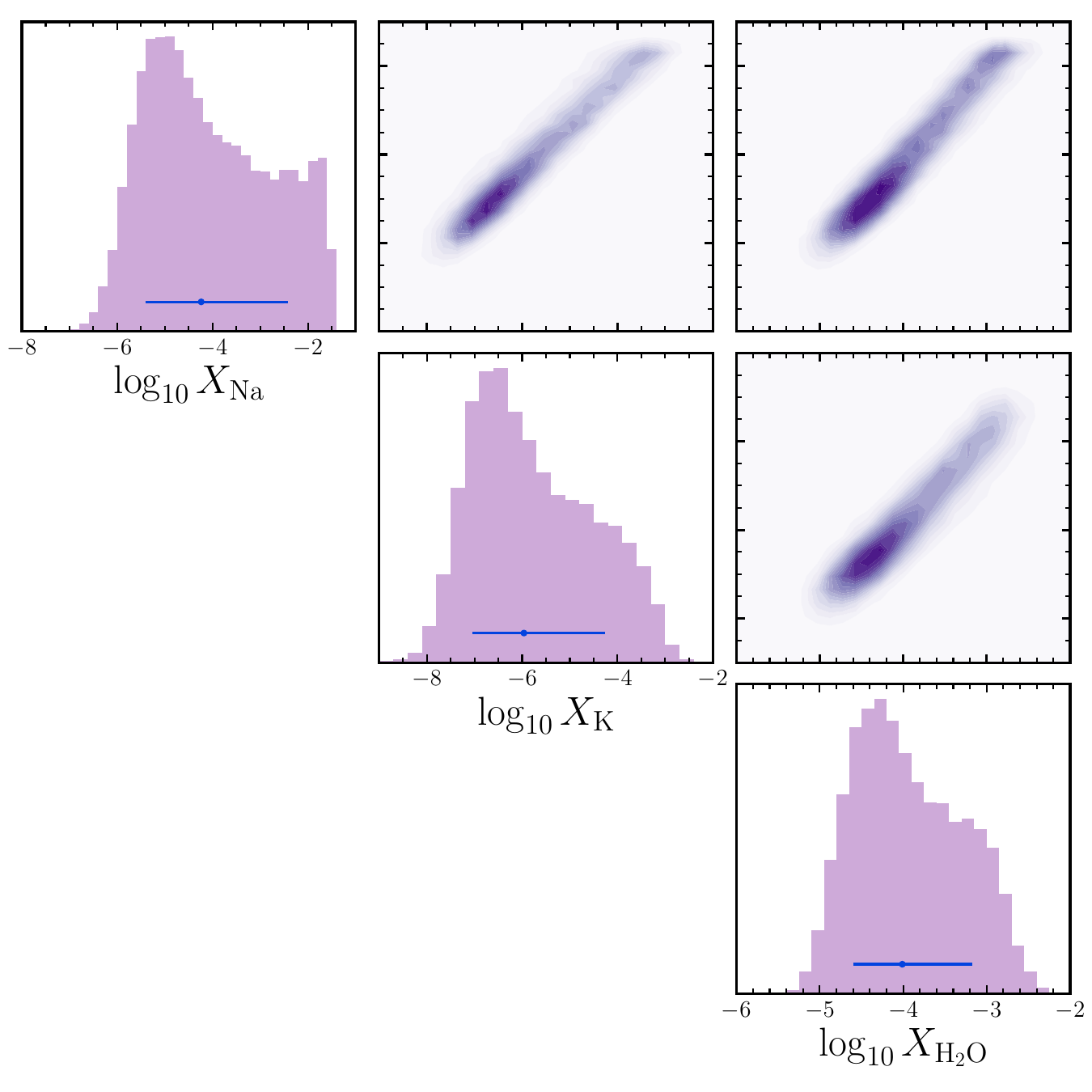}
\includegraphics[width=0.45\linewidth]{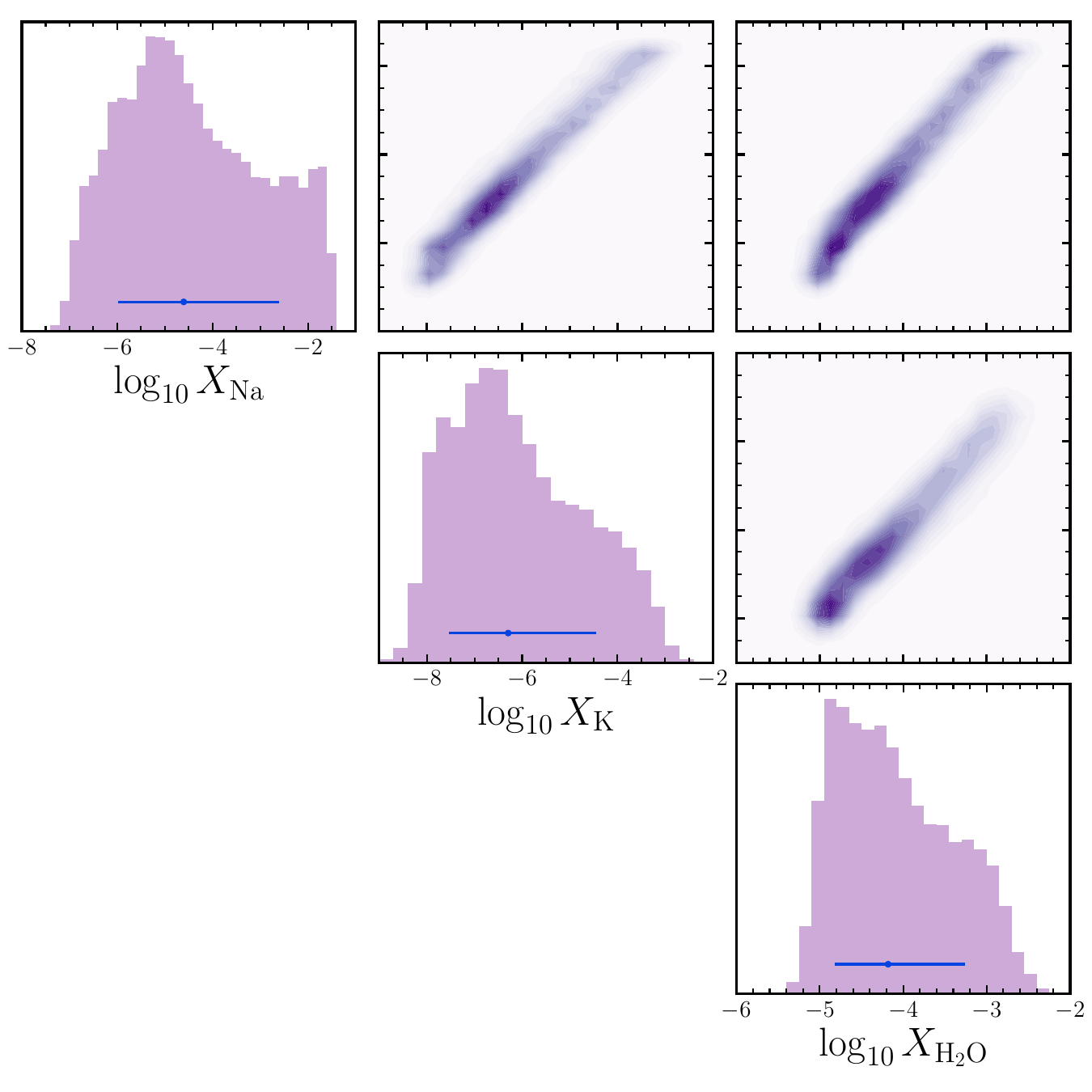}

	\def\arraystretch{1.0}
	\setlength{\tabcolsep}{5pt}
	\setlength{\arrayrulewidth}{1.3pt}
	\vspace{-11.0cm}\hspace{5.0cm}\begin{tabular}{cc}
        \multicolumn{2}{c}{\textbf{BMA}} \\
		\hline
		Parameter & Retrieved value \\
		\hline
		$\log_{10} X_{\rm Na}$ & $-4.57^{+1.86}_{-1.01}$ \\
		$\log_{10} X_{\rm K}$ & $-6.25^{+1.71}_{-0.93}$ \\
  		$\log_{10} X_{\rm H_2O}$ & $-4.16^{+0.86}_{-0.52}$ \\
		\hline
	\end{tabular}

	\def\arraystretch{1.0}
	\setlength{\tabcolsep}{5pt}
	\setlength{\arrayrulewidth}{1.3pt}
	\vspace{5.7cm}\hspace{-12.0cm}\begin{tabular}{cc}
        \multicolumn{2}{c}{\textbf{Pseudo BMA}} \\
		\hline
		Parameter & Retrieved value \\
		\hline
		$\log_{10} X_{\rm Na}$ & $-4.24^{+1.82}_{-1.17}$ \\
		$\log_{10} X_{\rm K}$ & $-5.96^{+1.71}_{-1.08}$ \\
  		$\log_{10} X_{\rm H_2O}$ & $-4.01^{+0.84}_{-0.59}$ \\
		\hline
	\end{tabular}

	\def\arraystretch{1.0}
	\setlength{\tabcolsep}{5pt}
	\setlength{\arrayrulewidth}{1.3pt}
	\vspace{-2.4cm}\hspace{5.0cm}\begin{tabular}{cc}
        \multicolumn{2}{c}{\textbf{Stacking}} \\
		\hline
		Parameter & Retrieved value \\
		\hline
		$\log_{10} X_{\rm Na}$ & $-4.61^{+2.00}_{-1.37}$ \\
		$\log_{10} X_{\rm K}$ & $-6.29^{+1.84}_{-1.24}$ \\
  		$\log_{10} X_{\rm H_2O}$ & $-4.18^{+0.93}_{-0.64}$ \\
		\hline
	\end{tabular}

    \vspace{0.5cm}\caption{Posterior distributions for the volume mixing ratios of three chemical species in the atmosphere of HD~209458b: H$_2$O, Na and K. These species are highlighted due to their prominent spectral features at the wavelengths of the observations. The upper left panel shows the posteriors from six of the seven candidate retrieval models, with no weighting applied. The remaining three panels show weighted average posteriors constructed using three methods: BMA (upper right), pseudo BMA (lower left) and stacking (lower right). Retrieved median values and 1-$\sigma$ uncertainties are shown as blue points with error bars and are provided in the three inset tables. The single-sector power-law haze + grey cloud model is omitted from this plot for clarity, since the posteriors strongly disagree with all other models, and the model is given no weight as a result of its poor performance.}
    \label{fig:3species}
\end{figure*}

Figure \ref{fig:3species} shows posterior distributions for the volume mixing ratios of three chemical species with strong absorption features within the wavelength range of the observations, Na, K and H$_2$O. We show the individual posteriors for all but one model as well as combined posteriors using each of the methods described in Section \ref{sec:methods}. We find that the cloud-free model as well as the two Ackerman \& Marley models find lower volume mixing ratios than the power-law haze + grey cloud models. When a grey cloud deck is imposed in the model, the size of the atmospheric column (the observable atmosphere, unocculted by clouds) is reduced, meaning that greater abundances of chemical species are required to explain the same spectral features \citep{Welbanks2019}. We find that the single-sector (i.e. fully cloudy) power-law + grey model is not capable of simultaneously fitting the optical slope along with the Na and K features, leading to a poor fit and retrieved abundances that are completely inconsistent with other models. As a result of its poor performance, this model is given no weight in any of the combined results, and so we omit these posteriors from Figure \ref{fig:3species} (though we still include the retrieved parameters for this model in Table \ref{tab:full_results} for reference).

For these three chemical species of interest, the retrieved parameters and associated uncertainties of the highest evidence model (4 sector power-law haze + grey cloud) are $\log_{10} X_{\rm Na} = -3.97^{+1.69}_{-1.25}, \log_{10} X_{\rm K} = -5.73^{+1.62}_{-1.15}$, and $\log_{10} X_{\rm H_2O} = -3.89^{+0.78}_{-0.63}$. If model selection by Bayesian evidence was used, these would be the final retrieved values, with the median result suggesting a sub-solar H$_2$O abundance, while remaining consistent with solar/super-solar expectations (i.e.,  $\log_{10} X_{\rm H_2O} \gtrsim-3.3$) within the 68\% confidence interval.

The retrieved values of each of these parameters following different model combination schemes are shown in Figure \ref{fig:3species}. Considering an ensemble of models using any of these methods leads to two main differences. First of all, the median estimates of the volume mixing ratios are decreased. This is a result of giving some weight to those models which are able to fit the absorption features with lower abundances, as described above. Second, the 1-$\sigma$ uncertainties found by model combination are either greater than, or similar to, the uncertainties found by model selection. The fact that a range of models shown here can produce reasonable fits to the data demonstrates that there is some model uncertainty, meaning we should expect our 1-$\sigma$ uncertainty estimates to increase when these models are included in the final posterior. By neglecting alternative candidate models and solely focusing on the model with the highest evidence, we are underreporting our uncertainty on the measured values of parameters of interest.

Furthermore, we note that in some cases, retrieved parameters are inconsistent with each other at a level $>$1-$\sigma$ between models (see Table \ref{tab:full_results}). If just a single model was selected from these candidates, then it could be possible to arrive at an answer which is both overconfident and incorrect. This highlights the importance of considering a range of model considerations in order to explore the full space of models that can explain a given data set. There is no way to obtain a ``ground truth'' via remote sensing of exoplanet atmospheres, meaning that all models will have shortcomings when trying to describe the real atmosphere that produced our observations. Conducting a hyper-marginalisation over all reasonable models using the methods we describe here should avoid the risk of a precise but inaccurate solution, provided enough models are considered and enough complexity is included in the models to capture the components of the actual atmosphere that are most relevant to our observations.

We can see from Figure \ref{fig:3species} that stacking leads to higher uncertainties on retrieved parameters than BMA and pseudo BMA. We can understand this by considering the weights given to each model according to each combination method (see Table \ref{tab:z_elpd}). BMA and pseudo BMA give very little weight to the models which yield lower chemical abundances, i.e.\ the A\&M models and the cloud-free model, with $<$2\% weight given to those three models in total. In contrast, stacking gives these three models 15.31\% of the total weight. This can be explained by considering the problem of model expansion described in Section \ref{subsec:bma}. The power-law haze + grey cloud models all produce similar best-fit spectra, regardless of the number of sectors considered, yielding similar posterior distributions. Since three different power-law cloud + grey haze models are included, we are effectively placing a larger prior mass on the power-law cloud + grey haze models than on the Ackerman \& Marley or cloud-free models when we use (pseudo) BMA. This leads to the power-law haze + grey cloud models taking more of the overall weight.

\begin{figure*}[t]
\centering
\includegraphics[width=\linewidth,trim={0 0 0 0},clip]{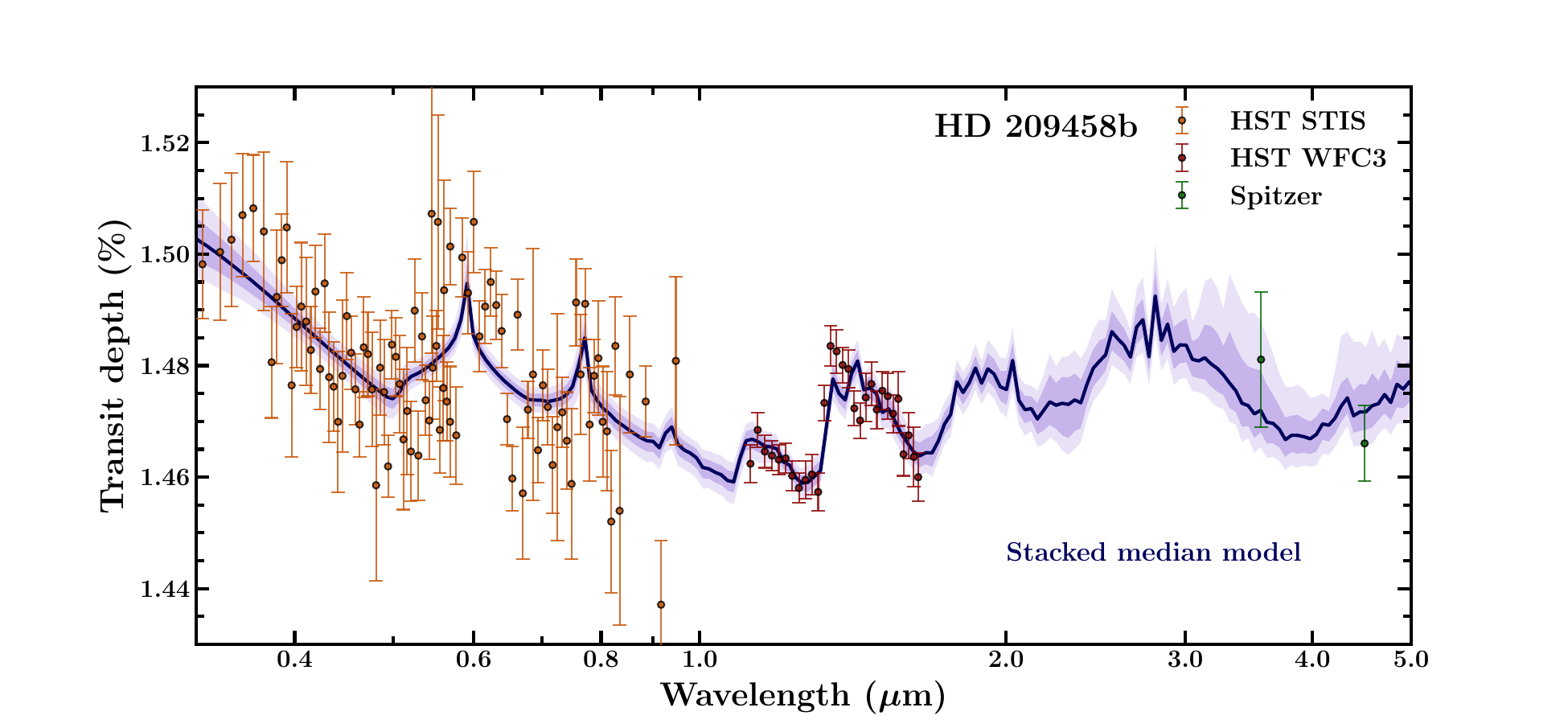}
    \caption{Transmission spectrum of HD~209458b, observed with HST STIS (orange points), HST WFC3 (red points), and Spitzer (green points). The purple line and shaded regions indicate the median model and 1- and 2-$\sigma$ confidence regions drawn from a weighted average posterior constructed by stacking, using the seven retrievals described in Section \ref{subsec:hd209}. All models are the same except for their treatment of clouds and hazes. Forward models have been downsampled to a resolution $R=200$ for clarity.}
    \label{fig:stacked_spectrum}
\end{figure*}

In contrast, stacking optimizes for the best distribution that can be constructed by combining the candidate model posteriors. This can be demonstrated by re-implementing the stacking with the highest-weighted candidate model removed (power-law + grey, four sectors). In this case, the model weight given to the three-sector power-law + grey increases drastically, to 58.72\%. Since the three-sector model has the closest posterior distribution to the four-sector model, stacking compensates for the loss of the four-sector model by assigning the three-sector model a much larger weight, in order to minimise the change in the resulting posterior. BMA and pseudo BMA do not have such flexibility, since the evidences and $\rm{elpd_{LOO}}$ scores for each model are fixed. Therefore, we find stacking to be our preferred method of combining the posterior distributions. We show the final stacked model spectrum in Figure \ref{fig:stacked_spectrum} and the stacked posterior distribution across all common parameters in Figure \ref{fig:stacked_post}.

\section{Summary and Discussion}
\label{sec:discussion}

\subsection{Recommendations for incorporating model uncertainty}

In this paper we have presented three ways to marginalise over model uncertainty by combining results from different candidate models: BMA, pseudo BMA and stacking of predictive distributions. These methods yield more realistic, and generally larger, uncertainties on retrieved parameters. Our case study in Section \ref{sec:results} demonstrates that both BMA and pseudo BMA are susceptible to bias from model expansion, where including a large number of models that are similar to each other could lead to those models being given an unreasonably high proportion of the total weight. Based on these findings, as well as other underlying properties of the three methods, we recommend stacking of predictive distributions to combine separately-fit posterior distributions to a given observation. However, we add that any of these approaches to incorporating model uncertainty are an improvement over model selection, which neglects model uncertainty entirely.

In the case study presented in Section \ref{sec:results}, several of the models can be considered as special cases of other models: for example, the three-sector power-law haze + grey cloud model is equivalent to the four-sector power-law haze + grey cloud model with $\phi_{c+h}=0$. The model performance metrics used in this work all aim to incorporate Occam’s razor in some manner, penalising excessive model complexity even if the quality of the fit to the data might be marginally better for a more complex model. Bayesian evidences do so by summing likelihoods over the whole parameter space; for a particularly complex model with many free parameters, the model may fit the data very poorly over a large region of the prior space, so the overall probability of the data under the model is low. LOO-CV does so by estimating out-of-sample predictive performance; if the model has been overfit to the data by fine-tuning specific parameters, then that model will be less likely to be able to accurately fit unseen data (in this case, the left-out points in the spectrum). Therefore, we can combine a simpler model (e.g. three-sector clouds/hazes) with a more complex expansion of that model (e.g. four-sector clouds/hazes) to find a weighted average that balances model complexity with quality of fit. However, we still recommend that consideration is given to the choice of candidate models, particularly if using BMA or pseudo-BMA due to the model expansion bias described above.

\subsection{What does model averaging tell us about the chemical composition of HD~209458b?}

As one of the most studied hot Jupiters in the field, the atmosphere and chemical composition of HD~209458b has been the subject of multiple studies in the past two decades \citep[e.g.,][]{Charbonneau2002, Deming2005, Madhusudhan2014, Line2016_hd209, Sing2016, MacDonald2017, Brogi2017, Welbanks2019_mz, Pinhas2019, Tsiaras2018, Barstow2017, Giacobbe2021}, with many of these studies focusing on its inferred H$_2$O abundance and associated atmospheric metallicity. Here, our most robust estimate resulting from the stacking approach results in a sub-solar H$_2$O abundance consistent with solar and super-solar values within 1-$\sigma$. 

However, we note that the resulting posterior distribution, while constrained, matches the expected bounds on the measured H$_2$O abundance for a cloud-free atmosphere \citep[e.g.,][]{Line2016}. The lower bound on the H$_2$O volume mixing ratio (i.e., $\sim$$10^{-6}$) corresponds to the limit where the abundance of the gas is low and the spectrum begins to be muted by the H$_2$-He CIA continuum. On the other hand, the upper bound on the H$_2$O volume mixing ratio (i.e., $\sim$$10^{-2}$) corresponds to the limit where the high abundance increases the mean molecular mass of the atmosphere, therefore decreasing the scale height of the planet, resulting in muted features. These effects have been all previously discussed in the literature, as well as possible ways to overcome them \citep[e.g.,][]{Benneke2012, Line2016, Welbanks2019, Welbanks2021_2D}.

Therefore, our results suggest that existing observations with HST may not be able to provide robust constraints on the planetary atmospheric abundances. The low signal-to-noise of these space-based observations, alongside the limited wavelength coverage, may not provide sufficient information to break these degeneracies alone. The presence of data in the optical (e.g., HST-STIS/HST-UVIS) has been suggested to provide additional information to break these degeneracies \citep[e.g.,][]{Welbanks2019} due to the presence of scattering slopes providing additional information \citep[e.g.,][]{Benneke2012}. However, for the particular case of HD~209458b and the models considered here, the information about this slope and its power to break this well-established degeneracy is dependent on the cloud/haze prescriptions employed. That is, previous abundance constraints using HST data only are strongly model dependent, as shown in Section \ref{sec:results}.

Looking towards the future, JWST observations of HD~209458b will likely be able to break this degeneracy and provide a more precise constraint on the planet's H$_2$O abundance, inferred metallicity and carbon-to-oxygen ratio. The presence of multiple spectral features due to absorption by numerous chemical species will make it possible to derive the relative abundances of the absorbing gases and as well as their absolute abundances \citep[e.g.,][]{Benneke2012}. Nonetheless, these infrared wavelengths will be affected by other model considerations like multidimensional effects, inhomogeneous clouds/hazes, night-side emission, among others \citep[e.g.,][]{MacDonald2017,Nixon2022}. Hence, in the presence of multiple models with different functionalities analyzing these data, we recommend implementing the model averaging techniques presented in this work in future studies of existing or newly observed data. 

\subsection{Practical considerations for implementing model combination}

The model combination methods described in this work do not add substantial computational cost, beyond the cost of running retrievals for each of the candidate models being considered. Assuming that Nested Sampling is used to obtain the posterior, no additional computations are required to obtain the Bayesian evidences for each model for use in BMA, making it the most straightforward of the three methods to implement. Even in the case where a different sampling algorithm is used that does not calculate the evidence, it may be possible to approximate it using, for example, the Bayesian Information Criterion, as employed in \citet{Wakeford2017}. 

Both pseudo BMA and stacking require more computational effort than BMA, since they require LOO-CV calculations for each candidate model. Thanks to the PSIS approximation, LOO-CV does not require $N_{\rm obs}$ model refits, since $\rm{elpd_{LOO}}$ scores can be estimated from the fit to the full data set. We note however, that the PSIS approximation may break down for certain data points, requiring the full refit with that data point left out \citep[e.g.,][]{Welbanks2023}. The need for a refit is indicated by the Pareto $k$ parameter, with $k>0.7$ indicating that the PSIS approximation failed for that point. Across all seven models considered, it was necessary to run additional retrievals for 5 data points in total out of a possible 868 ($N_{\rm obs} \times K$ models). While re-running retrievals on this data set was not particularly computationally costly, this could prove expensive for very large data sets with many more free parameters that might be needed for JWST, particularly if, for example, three-dimensional temperature parameterisations are required \cite[e.g,][]{Lacy2020,Welbanks2021_2D,Nixon2022}. In such cases, standard BMA may prove to be a more practical option, and would remain an improvement over model selection, provided that the various models are different enough from each other to avoid model expansion issues.

\begin{figure}
\centering
\includegraphics[width=\columnwidth,trim={0 0 0 0},clip]{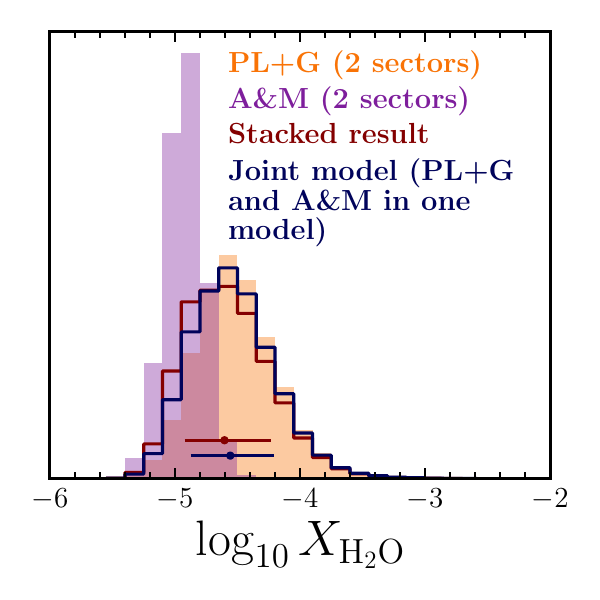}
    \caption{Retrieved volume mixing ratio of H$_2$O for the two-sector models, alongside their stacked posterior and the posterior for a model containing both aerosol prescriptions at once. We find that in this case, stacking yields a very similar posterior to a single model including both prescriptions.}
    \label{fig:am_plg}
\end{figure}

Another approach that one may consider when trying to account for several different model prescriptions could be to simply include all of the possible effects in a single model. In order to compare this to our model averaging methods, we show an additional retrieval in which both the power-law haze + grey cloud and the Ackerman \& Marley cloud are implemented in a single model. We use a two-sector model where all of the aerosol opacity is included in one of the sectors. We compare the result of this retrieval to the result of stacking the two-sector power-law haze + grey cloud and two-sector Ackerman \& Marley result in Figure \ref{fig:am_plg}. We find that the two approaches yield extremely similar posteriors, with both results remaining very close to power-law haze + grey cloud model. We note however that including all effects in one model may not always be equivalent to stacking, since the model including parameters from different prescriptions could introduce new degeneracies that may impact retrieved values of other parameters.

Although one could in principle try an approach where all possible modelling paradigms are considered in a single retrieval, we caution that this may lead to unphysical results or redundancy in the explored parameter space, and the results may be difficult to interpret. Furthermore, the high computational cost of such a model would likely be prohibitive. An extreme case of this would be to use a hierarchical Bayesian framework to simultaneously draw models from different models (even including models from different codes) to fit to a given spectrum. This would likely be the most comprehensive way to explore the space of candidate models, but we suggest model combination methods such as stacking as a more practical alternative.

\subsection{Applicability of model combination methods}

%How appropriate are these averaging methodologies when competing models have vastly different, inconsistent posteriors (e.g., parameter values that are many sigma's deviant from each other)? The text should discuss scenarios where these averaging methods might break down and/or not be appropriate to use.

%Please also discuss the applicability of the averaging methods for different model frameworks. For instance, some exoplanet atmosphere retrieval forward models assume some degree of self-consistency (e.g., grid-based fitting or chemical equilibrium along a temperature profile), in contrast to the free retrievals discussed in the manuscript. Notably, these self-consistent based methods fit directly for elemental abundance ratios (e.g., metallicity or C/O). These quantities can be computed from free retrieval abundance constraints (e.g., water, CO, methane, etc.). How would one go about averaging over these metallicity-C/O constraints under these different modeling paradigms? Could it be done in the same way?

The model combination approaches shown here can, in principle, be used to combine any set of models which share common parameters and which have been used to fit the same data set. It is possible that competing models may yield vastly different posteriors that are inconsistent with each other at several $\sigma$. In this case, if one model significantly outperforms the others according to the chosen metric, then the weighted posterior would very closely resemble the posterior from the most successful model. However, if some or all models perform similarly well, this would lead to a strongly multimodal averaged posterior distribution, reflecting the fact that it is impossible to distinguish between the candidate models. In this case, we would strongly advise against reporting median retrieved values for such parameters, and instead suggest presenting the individual posteriors and highlighting the discrepancy between different models.

As noted in Section \ref{sec:methods}, one drawback of BMA is that the Bayesian evidence can depend strongly on the chosen prior. For this reason, BMA is more applicable in the case where each model has the same free parameters and the same priors, and may even be preferred due to its relative simplicity and low computational cost. This scenario could arise if multiple retrieval codes are used to analyse the same observations, where small deviations between forward models due to the different implementations of radiative transfer can lead to differing results \citep{Barstow2020_comparison}. In fact, it is becoming common practice to include a number of independent retrieval analyses when presenting derived properties from JWST observations \citep[e.g.,][]{Taylor2023,Kempton2023}. By marginalising over differences in retrieval codes, we can obtain a more accurate representation of the community's best estimates of key atmospheric parameters. This could be seen as analogous to the Delphi method of obtaining consensus opinion from experts, which is widely applied in other domains \citep{Hsu2019}. We also note that the inclusion or removal of unconstrained parameters does not strongly affect the Bayesian evidence \citep{Trotta2008}. Therefore, it may be safe to use BMA in some cases where different parameters are used, if the posteriors of said parameters are prior-dominated, as is often the case for, e.g., parameters describing the $P$--$T$ profile when fit to HST transmission spectra \citep{Tsiaras2018}.

The case study presented in Section \ref{sec:results} focuses only on ``free chemistry'' retrievals, in which each chemical species of interest is represented by a free parameter and does not depend on, say, the thermal structure or abundances of other species. Different modelling and retrieval frameworks exist which instead calculate molecular abundances using overall metallicity as well as elemental abundance ratios such as C/O, assuming chemical equilibrium, with some disequilibrium processes occasionally included \citep[e.g.,][]{Line2013,Zhang2019}. These models are often termed ``chemically consistent''. Some previous works have converted free chemistry retrieval posteriors to match the chemically consistent format \citep[e.g.,][]{Coulombe2023}. This would enable the methods described here to be used to combine free chemistry and chemically consistent retrieval results. While this approach could be useful to get an overall estimate of the uncertainty of various retrieved parameters, we note that any similarities or differences between the results for such different modelling paradigms are interesting to consider, meaning that in these cases it would be prudent to highlight the results from individual modelling frameworks as well as the combined results.

\subsection{Concluding remarks}

The model combination methods described here can be applied to any problem in which a given data set is being fit by a number of candidate models. One potentially beneficial application of model combination could be at the data reduction stage, since, as with retrievals, it is becoming commonplace to include a number of independent data reductions when presenting new JWST spectra \citep[e.g.,][]{Ahrer2023,Alderson2023,Feinstein2023,Rustamkulov2023}, with different reductions showing some deviations between resultant spectra. Stacking could prove particularly useful here, since several reduction pipelines share initial stages, meaning that standard BMA could be susceptible to model expansion issues. Our methods also have many potential uses in other areas of exoplanet modelling, such as combining different internal structure modelling paradigms for sub-Neptunes \citep[e.g.,][]{Nixon2021,Neil2022}.

As we enter an era in which high-quality data from JWST and the ELTs will give us far more insight into exoplanet atmospheres than has ever been possible, it is more important than ever to ensure that our analysis procedures are robust and that we deal with uncertainty in a statistically sound manner. By failing to consider effects such as model uncertainty, thereby reporting overconfident measurements, we risk undermining the credibility of our claims, particularly if said claims are ultimately refuted when further observations are required. This is of particular relevance looking to the future, when the possibility of searching for biosignature gases becomes more realistic. By thinking carefully about all sources of uncertainty, we can establish trust in the many exciting discoveries that undoubtedly lie ahead in the study of exoplanets.

\begin{table*}

\centering

\def\arraystretch{1.0}
\setlength{\tabcolsep}{5pt}
\setlength{\arrayrulewidth}{1.3pt}

\begin{tabular}{cccccccc}
	\hline
	Parameter & PL+G & PL+G (2s) & PL+G (3s) & PL+G (4s) & A\&M & A\&M (2s) & Cloud-free \\
	\hline
    $\log_{10} X_{\rm Na}$ & $-6.79^{+3.41}_{-3.45}$ & $-5.42^{+0.64}_{-0.50}$ & $-4.32^{+1.58}_{-1.02}$ & $-3.97^{+1.69}_{-1.25}$ & $-6.58^{+0.20}_{-0.21}$ & $-6.40^{+0.26}_{-0.27}$ & $-6.75^{+0.21}_{-0.22}$ \\ 
    $\log_{10} X_{\rm K}$ & $-6.38^{+3.77}_{-3.70}$ & $-6.96^{+0.62}_{-0.51}$ & $-6.04^{+1.50}_{-0.96}$ & $-5.73^{+1.62}_{-1.15}$ & $-7.83^{+0.25}_{-0.26}$ & $-7.63^{+0.33}_{-0.36}$ & $-7.94^{+0.25}_{-0.26}$ \\ 
    $\log_{10} X_{\rm H_2O}$ & $-0.53^{+0.09}_{-0.07}$ & $-4.53^{+0.34}_{-0.30}$ & $-4.06^{+0.75}_{-0.52}$ & $-3.89^{+0.78}_{-0.63}$ & $-5.00^{+0.14}_{-0.15}$ & $-4.92^{+0.16}_{-0.17}$ & $-4.98^{+0.14}_{-0.15}$ \\ 
    $\log_{10} X_{\rm CH_4}$ & $-6.77^{+3.50}_{-3.52}$ & $-8.61^{+2.11}_{-2.22}$ & $-8.27^{+2.28}_{-2.50}$ & $-8.20^{+2.27}_{-2.49}$ & $-7.77^{+1.72}_{-2.70}$ & $-8.31^{+2.01}_{-2.35}$ & $-6.28^{+0.61}_{-3.45}$ \\ 
    $\log_{10} X_{\rm NH_3}$ & $-6.96^{+3.31}_{-3.32}$ & $-6.64^{+0.87}_{-3.51}$ & $-6.53^{+1.23}_{-3.56}$ & $-6.56^{+1.34}_{-3.55}$ & $-6.06^{+0.16}_{-0.24}$ & $-7.75^{+1.65}_{-2.72}$ & $-5.93^{+0.13}_{-0.14}$ \\ 
    $\log_{10} X_{\rm HCN}$ & $-8.42^{+2.94}_{-2.43}$ & $-9.46^{+2.28}_{-1.69}$ & $-9.29^{+2.35}_{-1.82}$ & $-9.18^{+2.45}_{-1.92}$ & $-9.85^{+1.95}_{-1.44}$ & $-9.60^{+1.95}_{-1.57}$ & $-9.81^{+1.97}_{-1.50}$ \\ 
    $\log_{10} X_{\rm CO}$ & $-3.90^{+1.64}_{-2.48}$ & $-6.31^{+1.63}_{-1.82}$ & $-5.88^{+1.84}_{-2.00}$ & $-5.69^{+1.87}_{-2.02}$ & $-7.08^{+1.44}_{-1.55}$ & $-6.60^{+1.54}_{-1.72}$ & $-7.01^{+1.38}_{-1.67}$ \\ 
    $\log_{10} X_{\rm CO_2}$ & $-8.56^{+2.84}_{-2.30}$ & $-9.88^{+1.62}_{-1.44}$ & $-9.71^{+1.82}_{-1.52}$ & $-9.65^{+1.91}_{-1.58}$ & $-10.19^{+1.43}_{-1.19}$ & $-9.96^{+1.49}_{-1.33}$ & $-10.21^{+1.50}_{-1.24}$ \\ 
    $T_0$ & $1399.52^{+102.27}_{-144.23}$ & $977.26^{+257.66}_{-129.31}$ & $1076.52^{+278.86}_{-204.31}$ & $1069.07^{+271.33}_{-193.76}$ & $838.38^{+64.85}_{-28.77}$ & $1243.07^{+214.03}_{-301.83}$ & $810.83^{+14.79}_{-7.71}$ \\ 
    $\alpha_1$ & $1.14^{+0.57}_{-0.59}$ & $1.18^{+0.54}_{-0.52}$ & $1.11^{+0.59}_{-0.55}$ & $1.03^{+0.63}_{-0.53}$ & $1.54^{+0.31}_{-0.43}$ & $1.02^{+0.62}_{-0.51}$ & $1.74^{+0.19}_{-0.32}$ \\ 
    $\alpha_2$ & $1.07^{+0.61}_{-0.61}$ & $0.97^{+0.67}_{-0.58}$ & $1.00^{+0.65}_{-0.59}$ & $0.99^{+0.66}_{-0.60}$ & $1.24^{+0.51}_{-0.62}$ & $0.90^{+0.69}_{-0.56}$ & $1.47^{+0.38}_{-0.62}$ \\ 
    $\log_{10} P_1$ & $-1.64^{+1.70}_{-1.81}$ & $-1.59^{+1.69}_{-1.76}$ & $-1.63^{+1.72}_{-1.75}$ & $-1.58^{+1.69}_{-1.78}$ & $-1.54^{+1.63}_{-1.85}$ & $-1.55^{+1.64}_{-1.70}$ & $-1.84^{+2.01}_{-1.92}$ \\ 
    $\log_{10} P_2$ & $-4.05^{+1.88}_{-1.35}$ & $-4.11^{+1.90}_{-1.31}$ & $-4.10^{+1.88}_{-1.32}$ & $-4.09^{+1.94}_{-1.33}$ & $-3.98^{+1.81}_{-1.36}$ & $-4.11^{+1.87}_{-1.29}$ & $-3.92^{+1.82}_{-1.37}$ \\ 
    $\log_{10} P_3$ & $0.60^{+0.99}_{-1.40}$ & $0.64^{+0.96}_{-1.33}$ & $0.59^{+0.98}_{-1.37}$ & $0.61^{+0.97}_{-1.33}$ & $0.60^{+0.97}_{-1.34}$ & $0.58^{+0.97}_{-1.24}$ & $0.47^{+1.09}_{-1.67}$ \\ 
    $\log_{10} P_{\rm ref}$ & $-4.23^{+0.22}_{-0.16}$ & $-2.66^{+0.67}_{-0.62}$ & $-3.18^{+0.80}_{-0.71}$ & $-3.31^{+0.79}_{-0.65}$ & $-0.83^{+0.09}_{-0.10}$ & $-2.36^{+0.60}_{-0.60}$ & $-0.71^{+0.06}_{-0.07}$ \\ 
    $\log_{10} a$ & $7.56^{+1.09}_{-1.44}$ & $4.38^{+0.71}_{-0.99}$ & $2.90^{+0.91}_{-0.89}$ & $3.22^{+1.03}_{-1.11}$ &  - &  - & - \\ 
    $\gamma$ & $-10.88^{+3.68}_{-3.03}$ & $-14.20^{+4.51}_{-3.81}$ & $-16.58^{+3.21}_{-2.33}$ & $-16.08^{+3.29}_{-2.64}$ &  - &  - & - \\ 
    $\log_{10} P_c$ & $-0.85^{+1.90}_{-1.93}$ & $-4.44^{+0.81}_{-0.55}$ & $-4.64^{+1.00}_{-0.89}$ & $-4.74^{+1.02}_{-0.83}$ &  - &  - & - \\ 
    $\log_{10} P_{\rm base}$ &  - &  - &  - &  - & $-0.19^{+1.24}_{-0.99}$ & $-2.35^{+1.37}_{-1.23}$ & - \\ 
    $\log_{10} f_{\rm cond}$ &  - &  - &  - &  - & $-5.86^{+3.19}_{-1.93}$ & $-2.08^{+1.36}_{-1.68}$ & - \\ 
    $\log_{10} K_{zz}$ &  - &  - &  - &  - & $8.36^{+1.31}_{-1.16}$ & $8.23^{+1.10}_{-0.94}$ & - \\ 
    $f_{\rm sed}$ &  - &  - &  - &  - & $3.05^{+1.26}_{-1.25}$ & $1.93^{+1.31}_{-0.65}$ & - \\ 
    $\phi_c$ &  - &  - & $0.60^{+0.07}_{-0.11}$ & $0.34^{+0.17}_{-0.20}$ &  - & $0.57^{+0.06}_{-0.15}$ & - \\ 
    $\phi_h$ &  - &  - & $0.30^{+0.08}_{-0.07}$ & $0.27^{+0.10}_{-0.10}$ &  - &  - & - \\ 
    $\phi_{c+h}$ &  - & $0.54^{+0.09}_{-0.12}$ &  - & $0.24^{+0.20}_{-0.15}$ &  - &  - & - \\ 
	\hline
	\end{tabular}
    \caption{Retrieved median estimates and 1-$\sigma$ uncertainties for for all parameters from each of the candidate models considered in this work, applied to the transmission spectrum of HD~209458b. PL+G refers to power-law haze + grey cloud models, and A\&M refers to Ackerman \& Marley cloud models. 2s, 3s and 4s refer to the number of cloud sectors in the model. Entries with a ``-'' indicate that the parameter is not included in that model. Units are shown in Table \ref{tab:priors}.}
    \label{tab:full_results}
\end{table*}

\section*{Acknowledgements}
The authors thank the anonymous referee, whose comments improved the quality of this manuscript. Our work was conceptualized during the first EXOplanet Model Interrogation and New Techniques Sprint (EXOMINTS) held at Arizona State University in 2023. EXOMINTS and this project benefited from the 2022 Other Worlds Laboratory (OWL) Mini-grant funded by the Heising-Simons Foundation. We thank Jonathan Fortney and the OWL for their support of this work, and Mike Line for helpful discussion regarding the implementation of the Ackerman \& Marley models. The authors acknowledge the University of Maryland high-performance computing resources used to conduct the research presented in this paper.  L.W.~acknowledges support for this work provided by NASA through the NASA Hubble Fellowship grant \#HST-HF2-51496.001-A awarded by the Space Telescope Science Institute, which is operated by the Association of Universities for Research in Astronomy, Inc., for NASA, under contract NAS5-26555. This work was performed under the auspices of the U.S. Department of Energy by Lawrence Livermore National Laboratory under Contract DE-AC52-07NA27344. The document number is LLNL-JRNL-855474. This research has made use of the NASA Astrophysics Data System and the Python packages \textsc{numpy} \citep{Harris2020}, \textsc{scipy} \citep{Virtanen2020}, and \textsc{matplotlib} \citep{Hunter2007}.
\includegraphics[width=\linewidth,trim={0 0 0 0},clip]{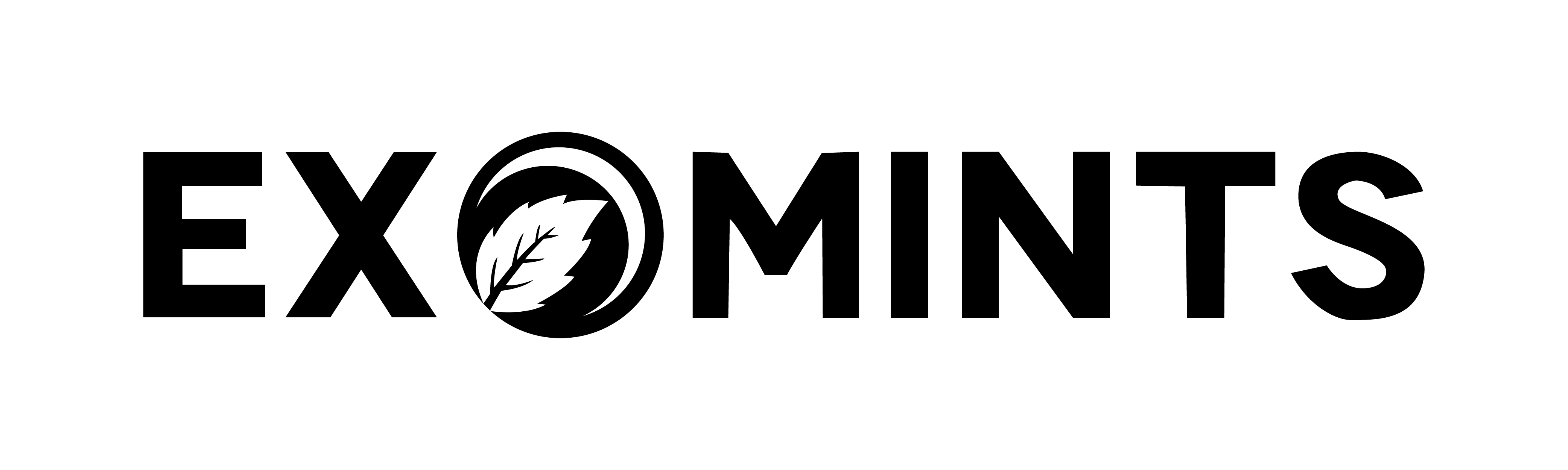}

\bibliographystyle{aasjournal}
\bibliography{main}

\appendix

\section*{Ackerman \& Marley cloud model}

Here we describe the method used to calculate extinction due to condensates using the Ackerman \& Marley cloud model \citep{Ackerman2001,Charnay2018}. First, the condensate mixing ratio $q_c$ is calculated as a function of pressure:
\begin{align}
    q_c = 
    \begin{cases}
    f_{\rm cond} \left( P/{P_{\rm base}} \right)^{f_{\rm sed}}, & P \leq P_{\rm base} \\
    0, & P > P_{\rm base}.
    \end{cases}
\end{align}
The nominal droplet size in each cloud layer, $r_{\rm sed}$, is given by
\begin{align}
    r_{\rm sed} = \frac{2\lambda}{3} \left( \sqrt{1+10.125\frac{\eta K_{zz}f_{\rm sed}}{gH_{\rm sc}(\rho_c-\rho_g)\lambda^2}}-1 \right),
\end{align}
where $g$ is the gravity, $\rho_c$ is the condensate particle density, and $\rho_g$ is the atmospheric density. The mean free path $\lambda$ is
\begin{align}
    \lambda = \frac{k_BT}{\sqrt{2} \, \pi d_B^2P},
\end{align}
where $k_B$ is Boltzmann's constant and $d_B=2.827^{-10}\,$m is the background gas molecule diameter \citep{Ackerman2001}. The dynamic viscosity of the background gas, $\eta$, is given by
\begin{align}
    \eta = \frac{125}{488} \frac{\sqrt{\pi \mu_g k_B T}}{\pi d^2} \left( \frac{k_BT}{\epsilon}\right)^{0.16},
\end{align}
where $\epsilon = 59.7k_B\,$K is the depth of the Lennard-Jones potential well for the atmosphere. We assume that the particle size follows a log-normal distribution, with geometric mean
\begin{align}
    r_g = r_{\rm sed} \exp \left( -\frac{\alpha+6}{2} \left( \log \sigma_{\rm eff} \right)^2 \right),
\end{align}
where $\sigma_{\rm eff} = 2$ is the log-normal distribution width. The total number concentration of particles $f_{\rm drop}$ is then
\begin{align}
    f_{\rm drop} = \frac{3 \mu_c q_c}{4 \pi \rho_c r_g^3} \exp \left( -\frac{9}{2} \left( \log \sigma_{\rm eff} \right)^2 \right).
\end{align}
In practice, the droplet mixing ratio is calculated for different radius bins of size $\Delta \log r$. As stated above, the particle size distribution is assumed to be log-normal, so we have
\begin{align}
    f_{\mathrm{drop}, r} &= f_{\rm drop} l(r) \Delta \log r, \\
    l(r) &= \frac{1}{\sqrt{2 \pi} \log \sigma_{\rm eff}} \exp \left( -\frac{1}{2} \left( \frac{r}{r_g \log \sigma_{\rm eff}} \right)^2 \right)
\end{align}
The extinction due to condensates at each particle radius is given by the product of the droplet mixing ratio and the Mie scattering opacity of the particle at that radius, with the total extinction given by the sum over particle radii. For the condensate species considered in this work, MgSiO$_3$, $\mu_c = 100.39\,$amu and $\rho_c = 3250\,$kg$\,$m$^{-3}$.

\begin{figure*}[t]
\centering
\includegraphics[width=\linewidth,trim={0 0 0 0},clip]{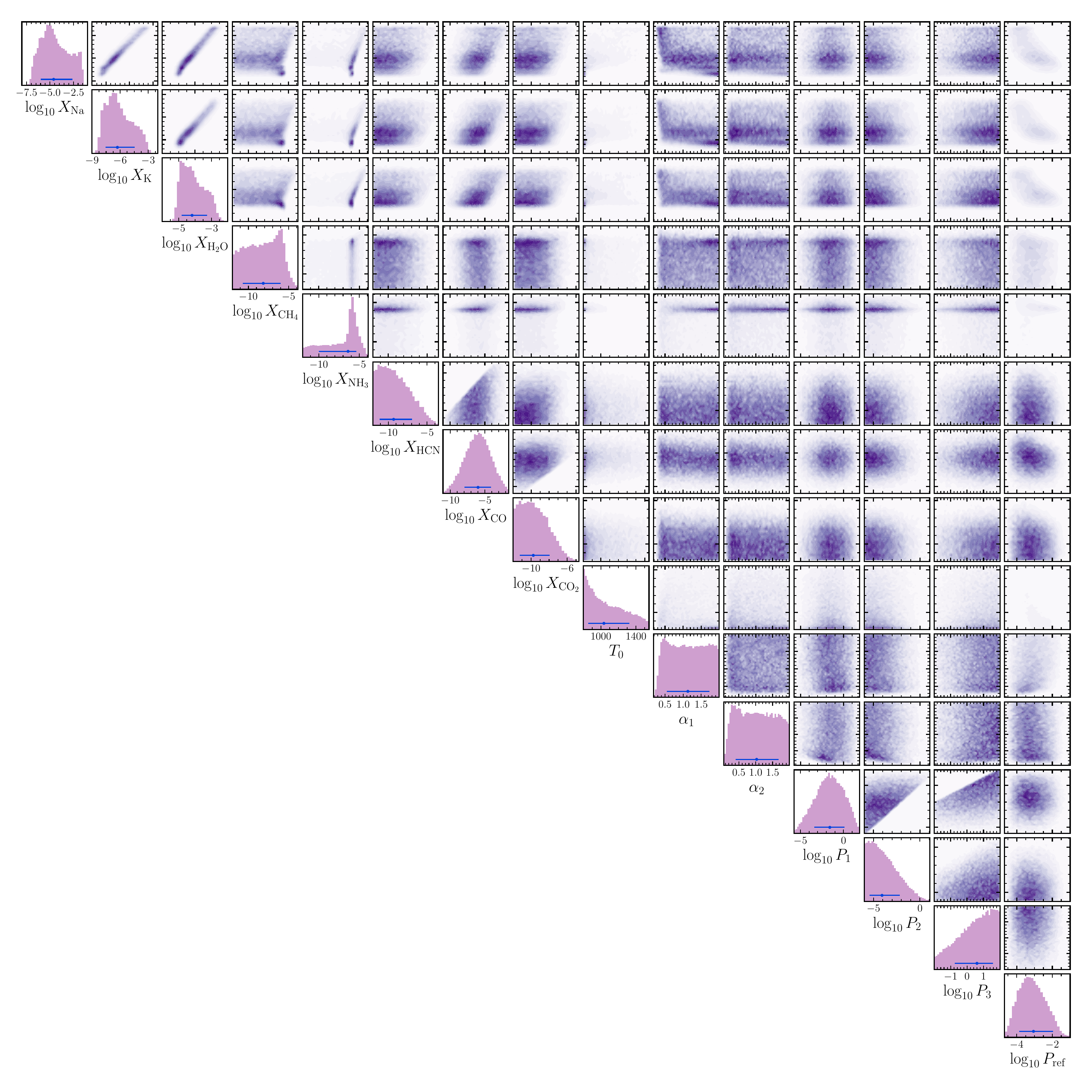}
    \caption{Stacked posterior distributions from retrievals of the transmission spectrum of HD~209458b for all shared parameters across the models considered in this work. Retrieved median values and 1-$\sigma$ uncertainties are shown as blue points with error bars and are provided in the inset table (units shown in Table \ref{tab:priors}).}
    \label{fig:stacked_post}

    \def\arraystretch{1.2}
	\setlength{\tabcolsep}{5pt}
	\setlength{\arrayrulewidth}{1.3pt}
	\vspace{-11.5cm}\hspace{-12.0cm}\begin{tabular}{cc}
		\hline
		Parameter & Retrieved value \\
		\hline
    $\log_{10} X_{\rm Na}$ & $-4.61^{+2.00}_{-1.37}$ \\ 
    $\log_{10} X_{\rm K}$ & $-6.29^{+1.84}_{-1.24}$ \\ 
    $\log_{10} X_{\rm H_2O}$ & $-4.18^{+0.93}_{-0.64}$  \\ 
    $\log_{10} X_{\rm CH_4}$ & $-8.15^{+2.19}_{-2.53}$  \\ 
    $\log_{10} X_{\rm NH_3}$ & $-6.39^{+1.02}_{-3.60}$  \\ 
    $\log_{10} X_{\rm HCN}$ & $-9.31^{+2.36}_{-1.82}$  \\ 
    $\log_{10} X_{\rm CO}$ & $-5.99^{+1.88}_{-1.96}$ \\ 
    $\log_{10} X_{\rm CO_2}$ & $-9.75^{+1.82}_{-1.51}$ \\ 
    $T_0$ & $1033.95^{+288.43}_{-180.24}$ \\ 
    $\alpha_1$ & $1.13^{+0.59}_{-0.58}$  \\ 
    $\alpha_2$ & $1.03^{+0.65}_{-0.62}$  \\ 
    $\log_{10} P_1$ & $-1.59^{+1.71}_{-1.79}$  \\ 
    $\log_{10} P_2$ & $-4.08^{+1.91}_{-1.33}$  \\ 
    $\log_{10} P_3$ & $0.60^{+0.98}_{-1.35}$ \\ 
    $\log_{10} P_{\rm ref}$ & $-3.06^{+1.10}_{-0.79}$ \\ 
		\hline
	\end{tabular}
 \vspace{3cm}
\end{figure*}

\end{document}